\documentclass[11pt]{article}

\usepackage[letterpaper]{geometry}

\usepackage{amsmath}
\usepackage{amsthm}
\usepackage{amsfonts}

\usepackage{mathptmx}
\usepackage{helvet}
\usepackage{newtxtext}
\usepackage{newtxmath}

\usepackage{microtype}

\usepackage{enumitem}
\setlist{noitemsep}

\theoremstyle{plain}
\newtheorem{theorem}{Theorem}

\theoremstyle{definition}

\usepackage[colorlinks,pagebackref,linkcolor=black,citecolor=black,urlcolor=black]{hyperref}

\usepackage{stmaryrd}                

\usepackage{bussproofs}
\EnableBpAbbreviations
\renewcommand{\fCenter}{\vdash}
\newcommand{\AXM}[1]{\AXC{$#1$}}
\newcommand{\UIM}[1]{\UIC{$#1$}}
\newcommand{\BIM}[1]{\BIC{$#1$}}

\newcommand{\AXN}[1]{\AX$#1$}
\newcommand{\BIN}[1]{\BI$#1$}
\newcommand{\UIN}[1]{\UI$#1$}
\newcommand{\TIN}[1]{\TI$#1$}
\newcommand{\QIN}[1]{\QuaternaryInf$#1$}

\newcommand\Forall[2]{\forall #1.\; #2}
\newcommand\Exists[2]{\exists #1.\; #2}

\newcommand{\ExistsUnique}[2]{\exists{!} #1.\; #2}
\newcommand{\PiT}[2]{\Pi #1.\; #2}
\newcommand{\SigmaT}[2]{\Sigma #1.\; #2}
\newcommand{\Lam}[2]{\lambda #1.\; #2}
\newcommand{\GenericBinder}[3]{#1 #2.\; #3}

\renewcommand\And{\mathrel\land}
\newcommand\Or{\mathrel\lor}
\newcommand\Implies{\rightarrow}
\newcommand\Iff{\leftrightarrow}
\newcommand\False{\bot}
\newcommand\True{\top}

\newcommand{\na}[1]{\mathrm{#1}}     
\newcommand{\fn}[1]{\mathsf{#1}}     
\newcommand{\mdl}[1]{\mathfrak{#1}}  
\newcommand{\ty}[1]{\mathsf{#1}}     

\newcommand{\Prop}{\ty{Prop}}
\newcommand{\Type}{\ty{Type}}

\newcommand{\bN}{\mathbb{N}}
\newcommand{\bZ}{\mathbb{Z}}
\newcommand{\bQ}{\mathbb{Q}}
\newcommand{\bR}{\mathbb{R}}

\newcommand{\proves}{\vdash}
\newcommand{\forces}{\Vdash}

\newcommand{\tval}[1]{\llbracket #1 \rrbracket}    

\title{Foundations\footnote{This is a draft of a chapter for the forthcoming \emph{Handbook of Proof Assistants and Their Applications in Mathematics and Computer Science}, edited by Jasmin Blanchette and Assia Mahboubi and published by Springer. Benno van den Berg and Lawrence C.~Paulson have served as second readers.}}

\author{Jeremy Avigad}
\date{\today}

\begin{document}

\maketitle

\noindent The goal of interactive theorem proving is to verify claims that are precise enough to be stated in mathematical terms. Even when it comes to reasoning about hardware and software, verification in a proof assistant amounts to writing a mathematical description of the system under consideration, writing a mathematical specification of the desired properties, and then proving that the system meets its specification.

The \emph{formal foundation} of a proof assistant is the formal deductive system with respect to which the correctness of the proof assistant's implementation is judged. The foundation specifies a language for defining objects and saying things about them, and it determines the means that can be used to prove that such statements are true. This chapter presents some general background in logic that is helpful for thinking and reasoning about formal systems, and it describes the three main families of foundations that are commonly used in proof assistants today.

\section{Overview}
\label{section:lf:overview}


\subsection{Design criteria}
\label{subsection:lf:design:criteria}

A fundamental requirement of any formal foundation is that it should be \emph{appropriate} to the task at hand. The underlying language should allow us to describe the objects we want to describe and say what we want to say about them, and we should be convinced that formal expressions in the system adequately represent our intended meaning. Moreover, we need to have faith that the formal methods of reasoning are sound and yield correct results. This is a matter of theology as much as science: we rely on our intuition, training, and respect for the history and culture of mathematics to support the judgment that our informal mathematical methods of reasoning are trustworthy, and we rely further on our intuition, and the shorter but still venerable tradition of formal logic, to support the judgment that the formal system adequately captures these forms of reasoning.

A formal foundation should also be \emph{simple}. The foundation is the abstract specification of the language and rules that are implemented by the system, and the more complicated the specification is, the harder it is to get the implementation right. The complexity also bears on the time it takes a proof assistant or verified checker to verify an individual proof.

Most proof assistants have a small, trusted computing base, the kernel, which is used to certify correctness. Some proof assistants additionally support recording and exporting proofs so that they can be checked by independent reference checkers. A number of ambitious projects have tried to increase confidence that the implementation of a foundation meets its abstract specification; one approach is to bootstrap the verification process by verifying the correctness of the kernel itself relative to an independent specification of the system or its semantics, sometimes even down to the level of machine code \cite{DBLP:conf/mkm/Carneiro20,DBLP:journals/jar/DavisM15,DBLP:conf/cade/Harrison06,DBLP:conf/popl/KumarMNO14,DBLP:journals/pacmpl/SozeauBFTW20}.

Finally, we also want the formal foundation to be \emph{expressive}. When we are working with a proof assistant, it is not enough to know that our theorems can be stated and proved in principle; we want it to be convenient to do so in practice, and that requires that interaction with the system be close to our natural forms of communication. We want it to be easy to define objects, state theorems, construct proofs, interpret error messages, find theorems in the library, and read and maintain code written by others. All these requirements bear on our choice.

In addition to the gap between the kernel implementation of a formal foundation and its abstract specification, there is also a gap between the user interface and the kernel. Users state and prove claims in an idealized language, much like a programming language, which has to be translated to the representations that are checked by the kernel. A \emph{parser} converts user input to internal representations, typically invoking an \emph{elaborator} to disambiguate expressions and infer information that can conveniently be left implicit.

There is a tension between the desire for simplicity and the desire for expressivity. Traditionally, foundational reduction has helped mediate between the two, allowing us to choose a simple foundation and then interpret a more expressive language and higher-level inferences on top of that. But then a proof assistant has to either transform the higher-level input to match the low-level specification, or trust a higher-level specification, which becomes the de facto foundation. So, although it is possible to narrow the gap between the user input and the standard of correctness by adopting a more elaborate foundation, there is an unavoidable tradeoff between expressivity and simplicity. The challenge, therefore, is to strike the right balance between the two.

\subsection{The landscape}
\label{subsection:lf:the:landscape}

To muddy the waters even further, the appropriate design criteria for a proof assistant depend on what we want to do with it. If we are interested in formalizing mathematics, our choice should depend on the type of mathematics we want to formalize, and if we are interested in verifying hardware or software, it should depend on the kinds of systems and properties we want to reason about. Some preferences even come down to a matter of taste, and arguments over the merits of different provers incite flame wars similar to those about the merits of different programming languages. This explains why there are so many different foundations on offer. The best we can do here is to map out the design landscape and to clarify some of the advantages and disadvantages of the various choices.

One paramount consideration is whether the foundation we choose is typed or untyped, and, if it is typed, how elaborate the type system is. In set theory, every object is a set, much the way that in (an idealized version of) the Lisp programming language, every object can be viewed as a list. In other words, in set theory, there is fundamentally only one type of object. In a typed foundation, as in a typed programming language, every expression has a type, and the type constrains what one can do with the expression. Having types makes it easier for a system to catch and flag errors, such as sending the wrong number of arguments to a function. It can also make user input more convenient, since information can be inferred from an expression's type. For example, in an expression $x + y$, if $x$ and $y$ are known to be integers, the system can infer the $+$ denotes integer addition. These benefits come at a price: expressions then \emph{have} to conform to the type discipline, constraining the user in ways that are sometimes unnatural.

Another consideration is whether the logical framework is classical or constructive. Constructive frameworks are designed to support a computational reading of definitions and theorems, whereby a proof of an existence statement comes either implicitly or explicitly with a construction of the object asserted to exist. Modern mathematics is thoroughly classical in that it allows us to define noncomputable functions and establish the existence of other noncomputable objects. But constructive foundations are often appealing to researchers in computer science, who prefer to think in computational terms.

Perhaps more to the point is the question as to how well the logical framework supports reasoning about computation. Constructive foundations often specify a notion of evaluation for terms in the language, which allows us to view terms as computer programs and reason about their behavior. But classical systems can also support extraction of code from sufficiently computable definitions, and there are various ways that provability in the system can be taken to establish properties of the extracted code. This enables one to reason about computational systems within a classical mathematical framework.

In this chapter, we will consider three families of foundations:
\begin{itemize}
\item set theory;
\item simple type theory;
\item dependent type theory.
\end{itemize}
Set theory was designed to provide a powerful and flexible foundation for mathematics, with axioms that are fairly simple and easy to understand. Set-theoretic foundations, implemented by systems like Mizar \cite{DBLP:journals/jfrea/GrabowskiKN10}, Isabelle/ZF \cite{MR2051585}, and Metamath \cite{DBLP:conf/tphol/Megill06}, are typically based on classical first-order logic.

The second family, simple type theory, descends from a tradition that began with Gottlob Frege, was further developed in the \emph{Principia Mathematica} of Alfred Whitehead and Bertrand Russell, and reached maturity in presentations by Alonzo Church. In the early 1980s, Michael J.~C.~Gordon, interested in hardware verification, inaugurated its use in proof assistants. It is typically implemented as a classical system and forms the basis for HOL4 \cite{DBLP:conf/tphol/SlindN08}, Isabelle/HOL \cite{DBLP:books/sp/NipkowPW02}, ProofPower \cite{DBLP:conf/tphol/Arthan01}, and HOL Light \cite{DBLP:conf/tphol/Harrison09a}.

The third family, dependent type theory, extends simple type theory with more elaborate means of defining types and expressions of those types. Such a type theory was the basis of N.~G. de~Bruijn's seminal \emph{Automath} proof assistant from the late 1960s. Most systems in use today descend more directly from systems of constructive mathematics introduced by Per Martin-L\"of, Jean-Yves Girard, and others in the 1970s. As we will see, the fact that they are based on functions and types as primitives brings them closer to the realm of programming languages, but there are also classical variations. Contemporary systems based on dependent types include Coq \cite{DBLP:series/txtcs/BertotC04}, PVS \cite{DBLP:conf/tphol/OwreS08}, Agda \cite{DBLP:conf/tphol/BoveDN09}, Nuprl \cite{DBLP:books/daglib/0068834}, Matita \cite{DBLP:conf/cade/AspertiRCT11}, and Lean \cite{DBLP:conf/cade/MouraKADR15}.

\subsection{A brief history of logic and foundations}
\label{subsection:lf:a:brief:history}

In the \emph{Prior Analytics}, Aristotle presented his theory of syllogisms, which are schematic rules of inference from two premises to a conclusion. The theory is based on the insight that some patterns of inference are \emph{valid}, meaning that every instance with true premises has a true conclusion. In other words, the validity of an inference does not depend on its contents, but, rather, its logical form. In this account, Aristotle provided a systematic characterization of the valid syllogistic forms.

Aristotle's logic was further developed in the scholastic philosophical tradition of the monastic schools and universities in the middle ages. In 1666, Leibniz proposed the development of a \emph{characteristica universalis}, a symbolic language for expressing propositions, and a \emph{calculus ratiocinator}, a calculus of reasoning. Most of his efforts in that regard, however, were not published or appreciated until the nineteenth century. A work by Antoine Arnauld and Pierre Nicole first published in 1662, \emph{La Logique ou l'Art de penser}, was more influential. Commonly known as the \emph{Port-Royal Logic} for the abbey near Paris where it was written, it incorporated but also critiqued the Aristotelian tradition, and combined it with the philosophical rationalism of Descartes. It also aimed to clarify the relationship between words, ideas, and objects and the nature of predication.

Logic became a part of mathematics in the nineteenth century with the work of George Boole and Augustus De Morgan, who viewed calculation with propositions as an instance of algebraic reasoning. Charles Peirce, Ernst Schr\"oder, and others extended the algebraic framework to include quantifiers. In his two-volume \emph{Grundgesetze der Mathematik} of 1893/1903, Gottlob Frege presented an axiomatic foundation for mathematics with functions, relations, quantifiers, and higher types. (In 1902, Russell showed that Frege's system is inconsistent, but eliminating the offending axiom results in essentially a version of higher-order logic.) Around the turn of the twentieth century, Giuseppe Peano also developed formal accounts of mathematical reasoning.

The nineteenth century also brought developments in the \emph{informal} foundations of mathematics. Augustin-Louis Cauchy, Georg Cantor, and Richard Dedekind all provided definitions of the real numbers in terms of sets or sequences of rationals, and Frege and Dedekind gave essentially set-theoretic constructions of the natural numbers.

Axiomatic foundations came of age in the twentieth century. Ernst Zermelo proposed the fundamental axioms of set theory in 1908, and in 1921 Abraham Fraenkel couched them in the framework of first-order logic and added the axiom of replacement. Russell and Whitehead presented a system of \emph{ramified type theory} in their landmark three-volume work, \emph{Principia Mathematica}, of 1910--1913. In 1940, Church introduced \emph{simple type theory}, incorporating ideas by Frank Ramsey and Leon Chwistek. The modern theories of constructive dependent type theory that are due to Per Martin-L\"of on the one hand and Thierry Coquand and G\'erard Huet on the other, discussed in Section~\ref{section:lf:dependently:typed:foundations}, owe a lot to the intuitionistic principles put forth by L.~E.~J.~Brouwer in the 1910s and formalized by Arend Heyting in the 1930s. They also draw on the \emph{Curry--Howard correspondence}, a relationship between logic and computation elucidated by Haskell Curry, William Howard, William Tait, and N.~G. de~Bruijn.

\section{First-Order Logic}

Philosophers often distinguish between a \emph{logical foundation} and a \emph{mathematical foundation}. The first is supposed to provide the general rules of reasoning in any domain, whereas the second is supposed to provide the rules for reasoning about mathematical objects in particular. First-order logic is an important foundation for logical reasoning in the first sense, since it is designed to model fundamental patterns of reasoning involving terms like ``and,'' ``or,'' ``not,'' ``every,'' and ``some.'' It can be interpreted as establishing patterns of inference that are valid in any domain that has at least one element.

\subsection{Syntax}
\label{subsection:lf:first:order:logic:syntax}

A first-order \emph{language} is a collection function symbols and relation symbols, each with an associated \emph{arity}, that is, a specification of the number of arguments. A $0$-ary function symbol is simply a constant symbol, whereas one can think of a $0$-ary relation symbol as standing for a fixed proposition. The syntax is then specified in two stages. Given a stock of variables $x, y, z, \dotsc$, the set of \emph{terms} is defined inductively, as follows:
\begin{itemize}
\item Each variable $x$ and constant symbol $c$ is a term.
\item If $f$ is an $n$-ary function symbol with $n \ge 1$ and $t_0, \ldots, t_{n-1}$ are terms, then $f(t_0, \ldots, t_{n-1})$ is a term.
\end{itemize}
The set of formulas is defined inductively, as follows:
\begin{itemize}
\item If $s$ and $t$ are terms, then $s = t$ is a formula.
\item If $R$ is an $n$-ary relation symbol and $t_0, \ldots, t_{n-1}$ are terms, then $R(t_0, \ldots, t_{n-1})$ is a formula.
\item $\bot$ is a formula, and if $A$ and $B$ are formulas, so are $A \And B$, $A \Or B$, and $A \Implies B$.
\item If $A$ is a formula and $x$ is a variable, then $\Forall x A$ and $\Exists x A$ are formulas.
\end{itemize}
Defining $\neg A$ to be $A \Implies \bot$, the set of propositional connectives chosen here is suitable for both classical and intuitionistic logic. For classical logic, smaller sets of connectives will do, and for both classical and intuitionistic logic, we can define $A \Iff B$ to be $(A \Implies B) \And (B \Implies A)$. The universal and existential quantifiers in the last clause are said to \emph{bind} the variable $x$. For example, $\Forall x {R(x, y)}$ says that everything is related to $y$, which is a statement about $y$ but not $x$. Intuitively, this formula says the same thing as $\Forall z {R(z, y)}$. Formulas that are the same up to a renaming of their bound variables are said to be \emph{$\alpha$-equivalent}, and many implementations use internal representations that identify $\alpha$-equivalent formulas (for example, using \emph{De Bruijn indices} \cite{DBLP:books/daglib/0067558}). Variables that are not bound are said to be \emph{free}, and a formula with no free variables is called a \emph{sentence}.


If $A$ is a first-order formula, $t$ is a term, and $x$ is a variable, then $A[t/x]$ denotes the result of substituting $t$ for $x$ in $A$. One has to be careful to rename bound variables when carrying out a substitution. For example, when interpreted as a statement about the natural numbers, $\Forall x {\Exists y} y > x$ says that for every number $x$, there is something larger than it. One should be able to replace $x$ by any expression, but naively replacing $x$ by $y + 1$ results in the statement $\Exists y {y > y + 1}$, which is patently false.

\subsection{Deductive systems}
\label{subsection:lf:first:order:logic:deductive:systems}

The following axiomatic deductive system was used by David Hilbert and his students in the 1920s:
\begin{enumerate}
\item $A \Implies (B \Implies A)$.
\item $(A \Implies (B \Implies C)) \Implies ((A \Implies B) \Implies (A \Implies C))$.
\item $A \Implies (B \Implies A \And B)$.
\item $A \And B \Implies A$.
\item $A \And B \Implies B$.
\item $A \Implies A \Or B$.
\item $B \Implies A \Or B$.
\item $(A \Implies C) \Implies ((B \Implies C) \Implies (A \Or B \Implies C))$.
\item $\bot \Implies A$.
\item $(\Forall x A) \Implies A[t/x]$.
\item $A[t/x] \Implies \Exists x A$.
\item $\Forall x {x = x}$.
\item $\Forall {x, y} {x = y \Implies t[x/z] = t[y/z]}$.
\item $\Forall {x, y} {x = y \Implies (A[x/z] \Implies A[y/z])}$.
\end{enumerate}
For classical logic, replace 9 with $\neg \neg A \Implies A$. The rules of inference are as follows:
\begin{enumerate}
\item From $A \Implies B$ and $A$ conclude $B$.
\item From $A \Implies B$ conclude $A \Implies \Forall x B$, provided $x$ is not free in $A$.
\item From $A \Implies B$ conclude $(\Exists x A) \Implies B$, provided $x$ is not free in $B$.
\end{enumerate}
The first is known as \emph{modus ponens}. If $\Gamma$ is a set of sentences of first-order logic, then $\Gamma \vdash A$, read ``$\Gamma$ proves $A$,'' expresses the statement that there is a formal proof of $A$ from hypotheses in $\Gamma$. The system of axioms above satisfies the \emph{deduction theorem}, namely, that if $\Gamma \cup \{ A \} \vdash B$ then $\Gamma \vdash A \Implies B$.

Many proof assistants implement systems of logical reasoning that are based on Gerhard Gentzen's systems of \emph{natural deduction}, which he first presented in the 1930s. In one formulation, formal derivations are taken to establish not formulas but \emph{sequents} of the form $\Gamma \vdash A$, where $\Gamma$ is a finite set of formulas. As the notation suggests, the sequent $\Gamma \vdash A$ is supposed to indicate that $A$ follows from $\Gamma$. In this context, it is convenient to write $\Gamma{,}\; A$ for $\Gamma \cup \{ A \}$ and to write $\Gamma{,}\; \Delta$ for $\Gamma \cup \Delta$. A novel feature of the natural deduction framework is that hypotheses can be introduced temporarily in a derivation. For example, the rule for showing that an implication $A \Implies B$ follows from $\Gamma$ is as follows: if $\Gamma{,}\; A \vdash B$ then $\Gamma \vdash A \to B$. In other words, the deduction theorem is incorporated as a basic rule of the calculus. The full set of rules is given in Figure~\ref{fig:natural:deduction}.
\setlength{\tabcolsep}{12pt}
\begin{figure}
\begin{center}
\begin{tabular}{cc}
\multicolumn{2}{c}{
\AXM{\strut}
\UIN{\strut\Gamma{,}\; A \fCenter A}
\DP
}
\\
\\
\AXN{\strut\Gamma \fCenter A}
\AXN{\strut\Delta \fCenter B}
\BIN{\strut\Gamma{,}\; \Delta \fCenter A \And B}
\DP
&
\AXN{\strut\Gamma \fCenter A \And B}
\UIN{\strut\Gamma \fCenter A}
\DP
\quad
\AXN{\strut\Gamma \fCenter A \And B}
\UIN{\strut\Gamma \fCenter B}
\DP
\\
\\
\AXN{\strut\Gamma \fCenter A}
\UIN{\strut\Gamma \fCenter A \Or B}
\DP
\quad
\AXN{\strut\Gamma \fCenter B}
\UIN{\strut\Gamma \fCenter A \Or B}
\DP
&
\AXN{\strut\Gamma \fCenter A \Or B}
\AXN{\strut\Delta, A \fCenter C}
\AXN{\strut\Pi, B \fCenter C}
\TIN{\strut\Gamma{,}\; \Delta{,}\; \Pi \fCenter C}
\DP
\\
\\
\AXN{\strut\Gamma{,}\; A \fCenter B}
\UIN{\strut\Gamma \fCenter A \Implies B}
\DP
&
\AXN{\strut\Gamma \fCenter A \Implies B}
\AXN{\strut\Delta \fCenter A}
\BIN{\strut\Gamma{,}\; \Delta \fCenter B}
\DP
\\
\\
\multicolumn{2}{c}{
\AXN{\strut\Gamma \fCenter \bot}
\UIN{\strut\Gamma \fCenter A}
\DP
}
\\
\\
\AXN{\strut\Gamma \fCenter A}
\UIN{\strut\Gamma \fCenter \Forall x A}
\DP
&
\AXN{\strut\Gamma \fCenter \Forall x A}
\UIN{\strut\Gamma \fCenter A[t/x]}
\DP
\\
\\
\AXN{\strut\Gamma \fCenter A[t/x]}
\UIN{\strut\Gamma \fCenter \Exists x A}
\DP
&
\AXN{\strut\Gamma \fCenter \Exists x A}
\AXN{\strut\Delta, A \fCenter B}
\BIN{\strut\Gamma{,}\; \Delta \fCenter B}
\DP
\\
\\
\multicolumn{2}{c}{
\AXM{\strut}
\UIN{\strut\Gamma \fCenter t = t}
\DP
}
\\
\\
\AXN{\strut\Gamma \fCenter r = s}
\UIN{\strut\Gamma \fCenter t[r/x] = t[s/x]}
\DP
&
\AXN{\strut\Gamma \fCenter r = s}
\AXN{\strut\Delta \fCenter A[r/x]}
\BIN{\strut\Gamma{,}\; \Delta \fCenter A[s/x]}
\DP
\end{tabular}
\end{center}
\caption{Natural Deduction}
\label{fig:natural:deduction}
\end{figure}
For classical logic, the rule {\em ex falso sequitur quodlibet} (from $\bot$ conclude anything) should be strengthened to {\em reductio ad absurdum}:
\begin{prooftree}
\AXN{\strut\Gamma{,}\; \lnot A \fCenter \bot}
\UIN{\strut\Gamma \fCenter A}
\end{prooftree}

Natural deduction satisfies the \emph{weakening} property, which is to say, if $\Gamma \fCenter A$ and $\Delta \supseteq \Gamma$ then $\Delta \fCenter A$. As a result, it is possible to present the rules in such a way that sets of hypotheses are preserved when reading a proof upwards. In this style, the introduction rule for conjunction would be written as follows:\strut
\begin{center}
  \AXN{\strut\Gamma \fCenter A}
  \AXN{\strut\Gamma \fCenter B}
  \BIN{\strut\Gamma \fCenter A \And B}
  \DP
\end{center}
This manner of description is better suited to proof search: it says that if you are looking for a proof of $A \And B$ from $\Gamma$, it suffices to find a proof of $A$ and $B$ from $\Gamma$. We will also use this style of presentation to describe typing judgments in Section~\ref{section:lf:dependently:typed:foundations}.

A set of sentences closed under provable consequences is called a \emph{theory}. So, if $T$ is a theory, $T \proves A$ is equivalent to $A \in T$. If $\Gamma$ is set of sentences, the set $\{ A \mid \Gamma \vdash A \}$ is a theory, called the \emph{theory axiomatized by $\Gamma$}.

It is often useful to use the vector notation $\vec x$ to denote a tuple of variables and similarly for a tuple of terms $\vec t$, and we will use this notation below.

\subsection{Sorts}
\label{subsection:lf:sorts}

In first-order logic, quantifiers range over a single, monolithic universe of objects. But we can \emph{relativize} a quantifier to a predicate, $A(x)$, as follows: $\Forall x {A(x) \Implies B(x)}$ says that every $A$ is a $B$, and $\Exists x {A(x) \And B(x)}$ says that some $A$ is a $B$. In a language with an order relation, the notation $\Forall {x < y} B(x)$ and $\Exists {x < y} B(x)$ are commonly used to abbreviate $\Forall x x < y \Implies B(x)$ and $\Exists x x < y \And B(x)$, respectively. Similarly, $\Forall {x \in y} A(x)$, $\Forall {x \subseteq y} A(x)$, $\Exists {x \in y} A(x)$, and $\Exists {x \subseteq y} A(x)$ are commonly used with the language of set theory.

In a slight variant of first-order logic known as \emph{many-sorted logic}, one has quantifiers and variables ranging over different domains of objects, and the function symbols and relation symbols come with a specification of the sorts of objects they expect. For example, a language designed to reason about geometry may have variables $p, q, r, \ldots$ ranging over points, variables $L, M, N, \ldots$ ranging over lines, and a relation $\fn{on}(p, L)$ that expresses that point $p$ is on line $L$. The sentence $\Forall {p, q} \Exists L {\fn{on}(p, L) \And \fn{on}(q, L)}$ expresses that there is a line through any two points, and
\[
\Forall {p, q, L, M} {p \ne q \And \fn{on}(p, L) \And \fn{on}(q, L) \And \fn{on}(p, M) \And \fn{on}(q, M) \Implies L = M}
\]
expresses that there is at most one line between two distinct points. Many-sorted logic can easily be interpreted in first-order logic without sorts by introducing predicate symbols for the sorts and relativizing quantifiers. But keeping track of an object's sort is such a fundamentally important task that many implementations prefer to build it into the logical framework.

\subsection{Definitional extensions}
\label{subsection:lf:definitional:extensions}

Fundamental to any proof assistant is the ability to extend a language with \emph{definitions}. Given any formula $A(\vec x)$, one can introduce a new relation symbol $R(\vec x)$ and the axiom $\Forall {\vec x} {R(\vec x) \Iff A(\vec x)}$. This is a \emph{conservative extension} in the sense that the new symbol~$R$ can be translated away, namely, by replacing it everywhere by $A$.

The formula $\Exists y {B(y) \And \Forall z {B(z) \Implies z = y}}$, which asserts that there is a unique~$y$ satisfying $B(y)$, is often abbreviated $\ExistsUnique y {B(y)}$. If a first-order theory proves $\Forall {\vec x} {\ExistsUnique y {A(\vec x, y)}}$, we can introduce a new function symbol $f(\vec x)$ to denote \emph{the} $y$ such that $A(\vec x, y)$ holds, together with the axiom $\Forall {\vec x} {A(\vec x, f(\vec x))}$. The formula $A(\vec x, y)$ provides what is known as a \emph{definite description} of the function $f$. The result is once again a conservative extension, and there is a tedious though straightforward way to eliminate the use of $f$ in favor of its definite description.

Extending a first-order theory with relations and functions defined in this way results in a \emph{definitional extension} of the original theory. If one drops the uniqueness requirement in the definition of a definite description and requires only that the theory proves $\Forall {\vec x} {\Exists y {A(\vec x, y)}}$, the formula $A$ is known as an \emph{indefinite description}. It is an important fact that with classical logic, adding a function symbol $f(\vec x)$ satisfying $\Forall {\vec x} {A(\vec x, f(\vec x))}$ still results in a conservative extension. The same is true with intuitionistic logic if we add the axiom of \emph{decidable equality}, namely, $\Forall {x, y} {x = y \Or x \ne y}$.

In classical logic, one can use an indefinite description to specify a function $f(\vec x)$ that returns a value $y$ satisfying $B(\vec x, y)$ if such a value exists. Such a function is known as a \emph{Skolem function}. In the presence of such a function, the formula $\Exists y {B(\vec x, y)}$ is equivalent to $B(\vec x, f(\vec x))$, so a Skolem function can be used to replace an existential quantifier. For that reason, Skolem functions are commonly used in automated reasoning.

\subsection{Partial functions and undefined terms}
\label{subsection:lf:partial:functions:and:undefined:terms}

In first-order logic, every function is assumed to return a value. But we often want to reason about partial functions: division by zero on the real numbers is undefined, a partial computable function may not return a value, the limit of a sequence may not exist, and a derivative may not exist at a point. In all three of the families of formal foundations we discuss below, as in first-order logic, every term is assumed to denote an object. Various options are then available for modeling a partial function:
\begin{enumerate}
\item Replace the partial function by a total version by specifying values where they would otherwise be undefined. For example, in many calculations, it is convenient to define $x / 0$ to be $0$ for every real number $x$. This device does not change the meaning of any statement that does not mention the division symbol; it merely introduces an alternative to division with similar properties.
\item Replace the partial function by a total version but leave the new values unspecified. For example, foundational frameworks often provide the means to specify an arbitrary real value $a$ without making any additional assumptions about it. One can then define $x / 0$ to be $a$.
\item Replace the partial function by a total function that returns a special value, $\bot$, interpreted as the value \emph{undefined}. With this approach, division on the reals is represented as a function from $\bR \times \bR$ to $\bR \cup \{ \bot \}$.
\item Replace the partial function by a total function on a smaller domain. With this approach, division on the reals is represented on a function from $\bR \times \{ x \in \bR \mid x \ne 0 \}$ to $\bR$.
\end{enumerate}
The last two approaches presuppose that the logical framework has mechanisms for reasoning about functions with various domains and codomains. For example, to adopt the third approach in type theory, for any data type $\alpha$, a new data type $\ty{Option} \, \alpha$ can be used to represent the extension of $\alpha$ by a new element to represent the value \emph{undefined}. But then in order to interpret an expression like $x / y + z$, the \emph{domains} of the operations of arithmetic also have to be extended to include undefined terms. We will see that dependent type theory provides helpful means of implementing the fourth approach, so that the expression $x / y$ comes with an explicit syntactic guarantee that $y$ is not $0$.

Implementing the third and fourth approaches in a logic of total objects is, in any case, challenging. One may reasonably maintain that definedness is so fundamental to mathematical and logical reasoning that it should be built into the logic rather than layered on top of it. Variations on first-order logic based on partial terms have been thoroughly studied in the literature \cite{MR786465,MR1396840,MR966421:modified}, and the IMPS proof assistant was based on a version of simple type theory with partial functions \cite{DBLP:journals/jar/FarmerGT93}.

\subsection{Semantics}
\label{subsection:lf:first:order:logic:semantics}

It is common to distinguish between a logical system's \emph{syntax} and its \emph{semantics}. A semantics for a logical system is an account of the meaning of its expressions and an indication as to what the formal rules of inference are supposed to accomplish. If we think of the syntax of a formal system as a specification of what it means for an implementation in a proof assistant to be correct, the semantics, in a sense, specifies what it means for the syntax to be correct.

In the classical, model-theoretic approach, terms in a system are supposed to denote objects in an intended interpretation and formulas are supposed to say things about an intended interpretation. The deductive apparatus is then supposed to establish the valid entailments, so that $\Gamma \proves A$ implies that $A$ is true in any interpretation that satisfies the assumptions in $\Gamma$. More precisely, a \emph{model} or \emph{structure} $\mdl M$ for a first-order language $L$ consists of a set of objects, $| \mdl M |$, called the \emph{universe} of the model, an interpretation of each $n$-ary function symbol of $L$ as an $n$-ary function on the universe of $\mdl M$, and an interpretation of each $n$-ary relation symbol as an $n$-ary relation on the universe of $\mdl M$. Given a term $t$ and an assignment $\sigma$ of values in the universe of $\mdl M$ to its free variables, we can recursively define the interpretation of $t$ in $\mdl M$, which we denote $\tval{t}^{\mdl M, \sigma}$. Similarly, given a formula $A$ and an assignment $\sigma$, we can specify what it means for $A$ to be \emph{true} in $\mdl M$ under that assignment, written $\mdl M \models_\sigma A$. If $\Gamma$ is a set of sentences, we say $\mdl M$ \emph{is a model of} $\Gamma$ if every sentence in $\Gamma$ is true in $\mdl M$. We say \emph{$\Gamma$ logically entails $A$} and write $\Gamma \models A$ if every model of $\Gamma$ is a model of $A$.

For first-order logic, we have the important fact that the syntactic and semantic notions of entailment coincide: $\Gamma \proves A$ if and only if $\Gamma \models A$. The forward direction, which says that the deductive rules of first-order logic make sense with respect to model-theoretic semantics, is known as \emph{soundness}. The reverse direction, which says that every logical entailment is formally derivable, is known as \emph{completeness}. It was first established by Kurt G\"odel.

According to the semantic definitions, $\mdl M \models A \Or \neg A$ for any model $\mdl M$ and sentence $A$. Developing a model-theoretic semantics for intuitionistic logic therefore requires more work, but it can be done. For example, a \emph{Kripke model} $\mdl K$ is a family of structures indexed by a partial order $P$, called \emph{worlds}, or \emph{nodes}. These can be thought of as representing partial states of information, and the definition specifies that if a node $\alpha$ is greater than a node $\beta$ in the partial order, then the universe of the structure at $\alpha$ includes the universe of the structure at $\beta$, and any atomic formula true at $\beta$ is also true at $\alpha$. The classical semantic relation $\mdl M \models_\sigma A$ is replaced by a relation $\mdl K, \alpha \forces_\sigma A$, which says that $A$ is true at node $\alpha$ in $\mdl K$ under assignment $\sigma$. Soundness and completeness for intuitionistic first-order logic now takes the following form: $\Gamma$ proves $A$ if and only if $A$ is true at any node of any Kripke model at which every sentence in $\Gamma$ is true. Soundness and completeness also hold with respect to a generalization of Kripke semantics known as \emph{Beth semantics}.

\emph{Algebraic semantics} generalizes model-theoretic semantics by assigning truth values in other algebraic structures. In classical first-order logic, formulas can be evaluated in any \emph{complete boolean algebra}, and in intuitionistic first-order logic, formulas can be evaluated in any \emph{complete Heyting algebra}. For intuitionistic logic especially, \emph{categorical logic} provides natural semantics as well. For example, intuitionistic first-order logic is sound and complete with respect to evaluation in any \emph{locally cartesian closed category} with finite coproducts, where sentences are interpreted as objects and a provable entailment $A \proves B$ gives rise to an arrow between the corresponding objects.

There are also computational approaches to the semantics of intuitionistic logic, and even classical logic. Such approaches are better at capturing the intuitive meaning of the intuitionistic connectives, and will be discussed in Section~\ref{subsection:lf:simply:typed:lambda:calculus:semantics}.

\subsection{Decidability and incompleteness}
\label{subsection:lf:decidability:and:incompleteness}

We will see in Section~\ref{section:lf:simply:typed:lambda:calculus} that logic provides a means to reason about computation, and that formal derivations can have computational content. Conversely, from a metatheoretic standpoint, we can carry out computations involving terms, formulas, and derivations. A first-order theory $T$ is \emph{decidable} if there is an algorithm that, on input $A$, determines whether or not $T$ proves $A$. As it turns out, some of the theories that come up in logic are decidable and some are not, and it is often an interesting challenge to determine which is the case.

We should not expect a reasonably expressive theory to be decidable. For example, imagine we have a formal language in which we can express the statement ``Turing machine $M$ halts when started on empty input.'' Suppose moreover that we have a theory, $T$, with the property that if $M$ halts on empty input then $T$ is strong enough to prove that fact, and, conversely, if $T$ proves that $M$ halts on empty input, then it does. Then deciding whether or not $M$ halts on empty input is tantamount to deciding whether or not $T$ proves that fact, so the halting problem is reducible to $T$.

If the theory $T$ just described has a computable set of axioms, then the halting problem is in fact equivalent to $T$: given a formula $A$, we can describe an algorithm that searches for a proof of $A$ from the axioms of $T$, and the question as to whether $A$ is in $T$ reduces to the question as to whether that Turing machine halts.

A theory $T$ is said to be \emph{consistent} if $T$ does not prove $\bot$, and \emph{complete} if, for every sentence $A$, either $T$ proves $A$ or $T$ proves $\neg A$. In other words, a theory is complete if it is strong enough to settle the truth of every sentence in its language. A complete, computably axiomatized first-order theory $T$ is necessarily decidable: if $T$ is inconsistent, then it proves everything and the decision procedure is trivial, and if $T$ is consistent, there is an algorithm that decides whether or not $T$ proves $A$ by systematically searching for a proof of $A$ or $\neg A$ from the axioms of $T$. Combining this observation with the one in the previous paragraph shows that no reasonably expressive theory of mathematics is complete.

G\"odel's \emph{first incompleteness theorem} strengthens this result and makes it precise. The following formulation relies on a finite set of axioms defined by Raphael Robinson, denoted $Q$, that embodies very modest assumptions about basic arithmetic on the natural numbers \cite{MR1640326}. The theorem applies to any theory in the language of arithmetic that includes $Q$. It applies more generally to any theory $T$ that \emph{interprets} $Q$ in the sense that it is possible to translate the language of arithmetic into the language of $T$ in such a way that the axioms of $Q$ are provable.
\begin{theorem}
There is no complete, consistent, computably axiomatized theory that interprets $Q$.
\end{theorem}
\noindent G\"odel's original proof (of a slightly weaker statement) involves constructing a sentence in any such theory that says ``I am not provable,'' but one can also arrive at the result via computability considerations similar to the ones above. The \emph{second incompleteness theorem}, which we will not state precisely here, says, roughly, that no reasonable theory of mathematics can prove its own consistency.

The undecidability result sketched above can also be strengthened. A theory $T$ in the language of arithmetic is undecidable if it is \emph{consistent} with $Q$, that is, $T \cup Q$ is consistent. More generally, a theory is undecidable if it is consistent with a translation of the axioms of $Q$ to its language. Since $Q$ can be interpreted in set theory and pure first-order logic is consistent with that interpretation, provability in first-order logic in a language with a single binary relation symbol is undecidable. The intuition behind this result is that the binary relation \emph{could be} the element-of relation in set theory, which means that a provability query can encode a question about natural numbers representing Turing machine computations.

This is worse than guilt by association; it is guilt by virtue of having the mere capacity to commit a crime. Imagine a police officer pulling theory $T$ over and saying ``Hey, you! You look like a theory of arithmetic!'' Theory $T$ says ``What? Me? A theory of arithmetic? I don't know anything about arithmetic!'' The police officer takes a good look at $T$ and says, ``Well, you look like you \emph{could} be a theory of arithmetic. Get outta here! You're undecidable!''

The fact that interesting mathematical theories are undecidable, however, provides job security for people working in automated and interactive theorem proving. It tells us, in a sense, that proof search and user input are unavoidable.

A number of important first-order theories are decidable. If $\mdl M$ is a model, the \emph{theory of $\mdl M$} is the set of sentences $\{ A \mid \mdl M \models A \}$. The following are all decidable:
\begin{enumerate}
\item The theory of the natural numbers, $\bN$, or the integers, $\bZ$, in a language with $0$, $1$, $+$, and $\le$, but without multiplication. These are known as \emph{Presburger arithmetic} or \emph{integer linear arithmetic}.
\item The theory of the real numbers, $\bR$, in the same language. This is known as \emph{linear arithmetic}. It can be axiomatized as the theory of a densely ordered abelian group (additively written) with $0 < 1$, and so the theory does not change if $\bR$ is replaced by any densely ordered subgroup, like $\bQ$.
\item The theory of the real numbers, $\bR$, in the same language together with multiplication. This is known as the theory of \emph{real closed fields}.
\item The theory of the complex numbers as a field. This is known as the theory of \emph{algebraically closed fields}.
\end{enumerate}
In addition, some useful fragments of first-order logic are decidable. For example, a sentence of first-order logic is said to be \emph{universal} if it can be written $\Forall {\vec x} A$, where $A$ is quantifier-free. An algorithm known as \emph{congruence closure} can be used to decide the universal fragment of first-order logic with equality.

Results like these form the basis for \emph{decision procedures}, and are useful in the context of interactive theorem proving because users can rely on them to prove statements within their scope automatically. In that context, we generally want a procedure that not only decides the truth of a sentence but also provides an explicit proof of the sentence when the answer is positive.

\subsection{Arithmetic}
\label{subsection:lf:arithmetic}

Theories of the natural numbers have held an important place in the logical tradition, not only because the natural numbers are so fundamental, but also because they can be used to \emph{code} or \emph{represent} other finite objects. As a result, a theory of the natural numbers can be seen as providing a foundation for finitary mathematics.

The set of \emph{primitive recursive functions} is the set of functions from $\bN$ to $\bN$, of various arities, defined inductively as follows:
\begin{itemize}
\item The constant $0$ (as a $0$-ary function) and the function $\fn{succ}(x) = x + 1$ are primitive recursive.
\item If $t(\vec x)$ is any term involving primitive recursive functions that have previously been defined, the function $f$ defined by $f(\vec x) = t(\vec x)$ is primitive recursive.
\item If $g(\vec y)$ and $h(x, z, \vec y)$ are primitive recursive, so is the function $f(x, \vec y)$ recursively defined by the following equations:
\begin{align*}
  f(0, \vec y) & = g(\vec y) \\
  f(x+1, \vec y) & = h(x, f(x, \vec y), \vec y)
\end{align*}
\end{itemize}
In words, the set of primitive recursive functions is the smallest set containing zero and successor and closed under \emph{explicit definition} and \emph{primitive recursion}. As an example of an explicit definition, if $g(x, y, z)$ and $h(x)$ have previously been defined, we can introduce a new primitive recursive function $f(x, y)$ with defining equation $f(x, y) = g(x, h(x), h(y))$. This clause can be replaced by a more rigid schema for composing functions if we include projections $p^n_i(x_0, \ldots, x_{n-1}) = x_i$ among the basic functions.

\emph{Primitive recursive arithmetic}, or $\na{PRA}$, is a first-order theory with a symbol for each (description of a) primitive recursive function. Its axioms consist of the defining equations for functions defined by explicit definition or primitive recursion, axioms that say that $\fn{succ}(x)$ is injective and zero is not the successor of induction, and a schema of induction for quantifier free formulas $A(x)$:
\[
  \Forall {\vec z} {A(0) \And (\Forall x {A(x) \Implies A(\fn{succ}(x))}) \Implies \Forall x A(x)}
\]
Here $\vec z$ denotes the free variables of $A(x)$ other than $x$. $\na{PRA}$ can also be presented as a quantifier-free calculus, or even as nothing more than a calculus of equations of the form $s = t$ between terms in the language. Some historians have argued that $\na{PRA}$ represents the informal notion of \emph{finitistic mathematics} put forth by David Hilbert in the early twentieth century \cite{tait:81}.

First-order arithmetic is essentially primitive recursive arithmetic together with the schema of induction for \emph{all} formulas in the language. More precisely, \emph{Heyting arithmetic}, or $\na{HA}$, is the version of this theory based on intuitionistic logic, and \emph{Peano arithmetic}, or $\na{PA}$, is the version based on classical logic. These theories are usually formulated in a smaller language consisting of $0$, $\fn{succ}$, $+$, $\times$, and $\le$. Using coding to represent sequences of numbers and using induction, one can then introduce the primitive recursive functions in a definitional extension.

By the first incompleteness theorem, $\na{HA}$, $\na{PA}$, and $\na{PRA}$ are undecidable. All of the mathematical foundations we will consider here support definition by primitive recursion and proof by induction, and most are vastly more powerful than these theories. But even $\na{PRA}$ turns out to be a surprisingly robust theory for finite mathematics. Remarkably, we do not know of any published theorems of mathematics that can naturally be expressed in $\na{PRA}$ but not proved there, short of theorems that were designed by logicians specifically to have that property \cite{MR2006194}.

The original Boyer--Moore theorem prover, \emph{NQTHM}, and its successor, \emph{ACL2}, are based on quantifier-free logics similar in spirit to primitive recursive arithmetic. ACL2 was later extended with a principle of induction up to the ordinal $\varepsilon_0$, which makes the system slightly stronger than first-order arithmetic. For details, see \cite{DBLP:books/daglib/0098268}.

Theories even weaker than primitive recursive arithmetic have been studied, especially in relation to computational complexity. See, for example, \cite{MR1640326}.

\section{Simply Typed Lambda Calculus}
\label{section:lf:simply:typed:lambda:calculus}

Just as first-order logic provides a framework for modeling logical reasoning, the simply typed lambda calculus provides a framework for reasoning about functions. It includes notions of \emph{lambda abstraction} and \emph{computational reduction} that are fundamental to a number of contemporary proof assistants.

\subsection{Syntax}
\label{subsection:lf:simply:typed:lambda:calculus:syntax}

The set of \emph{simple types} over a collection $\mathcal B$ of \emph{basic types} is defined inductively, as follows:
\begin{itemize}
\item Every basic type $B \in \mathcal B$ is a type.
\item If $\alpha$ and $\beta$ are types, so is $\alpha \to \beta$.
\end{itemize}
Types are just syntactic expressions, but the idea is that an object of type $\alpha \to \beta$ is a function from $\alpha$ to $\beta$. The simple types are also called the \emph{finite types} because they are built up using finitely many iterations of the arrow construction. Given a collection $\mathcal C$ of constant symbols, each with an assigned type, the set of terms of the simply typed lambda calculus and their associated types are defined inductively as follows:
\begin{itemize}
  \item Each constant $c \in \mathcal C$ is a term of its assigned type.
  \item There are infinitely many variables $x, y, z, \ldots$ of each type.
  \item If $t$ is a term of type $\beta$ and $x$ is a variable of type $\alpha$, the term $\Lam x t$ is a term of type $\alpha \to \beta$.
  \item If $t$ is a term of type $\alpha \to \beta$ and $s$ is a term of type $\alpha$, the term $t \, s$ is a term of type $\beta$.
\end{itemize}
In the intended interpretation, $\Lam x t$ denotes the function that maps $x$ to $t$, and $t \, s$ is the result of applying $t$ to $s$. The fact that a term $t$ has type $\alpha$ is often written $t : \alpha$.

For example, suppose we start with the basic type $\ty{Nat}$ and constants $0 : \ty{Nat}$ and $\fn{succ} : \ty{Nat} \to \ty{Nat}$. Then assuming the variable $x$ has type $\ty{Nat}$ and the variable $f$ has type $\ty{Nat} \to \ty{Nat}$, the term $\Lam f {\Lam x {f \, (\fn{succ} \, (\fn{succ} \, x))}}$ has type $(\ty{Nat} \to \ty{Nat}) \to (\ty{Nat} \to \ty{Nat})$. Intuitively, this term denotes the function that maps any function $f$ to the function $x \mapsto f(x + 2)$.

Just as a quantifier binds a variable in a first-order formula, lambda abstraction binds a variable in a lambda expression. Terms that are equivalent up to renaming bound variables are said to be $\alpha$-equivalent and are generally treated as syntactically identical. As with first-order logic, some care is needed to define the notion $t[s/x]$ of substituting $s$ for $x$ in $t$, renaming bound variables if necessary.

\subsection{Reduction}
\label{subsection:lf:simply:typed:lambda:calculus:reduction}

Terms in the simply typed lambda calculus can be \emph{reduced}. A term of the form $(\Lam x t) \, s$ is called a \emph{$\beta$-redex}, and it \emph{$\beta$-contracts} to $t[s/x]$. A term $s$ \emph{$\beta$-reduces to} a term $t$ if $t$ can be obtained from $s$ by iteratively contracting $\beta$-redexes that occur anywhere in the term. One can show that reduction does not change the type of a term, a property known as \emph{subject reduction}. A term that has no $\beta$-redexes is said to be \emph{irreducible} or in \emph{$\beta$-normal form}, and reducing a term to one in normal form is known as \emph{normalization}.

You can think of normalization as a notion of computation. In the simply typed lambda calculus with a basic type $\ty{Nat}$ and constants $0$ and $\fn{succ}$, the only closed terms of type $\ty{Nat}$ are numerals, that is, expressions of the form $\fn{succ} \, (\fn{succ} \, \cdots \, (\fn{succ} \, 0) \cdots )$. So, in that setting, reducing a term to normal form is tantamount to evaluating it to a natural number. The computational interpretation of normalization is bolstered by the fact that the simply typed lambda calculus is \emph{strongly normalizing} and \emph{confluent}. Strong normalization means that, starting from any term, there is no infinite sequence of nontrivial one-step reductions, so that any method of reducing a term eventually results in a normal form. Confluence means that if a term $s$ reduces to both $t$ and $u$, then $t$ and $u$ in turn can be reduced to a common term $v$. This implies that normal forms are unique.

$\beta$-reduction is often combined with other reductions. An expression $\Lam x t x$ is said to \emph{$\eta$-contract} to $t$ if $x$ is not free in $t$. Whereas $\beta$-contraction for lambda abstraction eliminates an abstraction followed by application, $\eta$-contraction works the other way around, eliminating an application followed by an abstraction. The $\eta$ rule is almost always considered together with $\beta$, giving rise to the notion of $\beta\eta$-reduction, $\beta\eta$-normal form, and so on. The simply typed lambda calculus is strongly normalizing and confluent with respect to $\beta\eta$-reduction as well.

\subsection{Equational theories}
\label{subsection:lf:simply:typed:lambda:calculus:equational:theories}

There are deductive systems for the simply typed lambda calculus associated with $\beta$- and $\beta\eta$-reduction, designed to establish equations $s = t$ between terms. In addition to rules expressing the reflexivity, symmetry, and transitivity of equality, we include the following two rules, which say that equality is preserved under lambda abstraction and application. In the first rule, $s$ and $t$ are assumed to have the same type, and in the second rule, $s$ and $t$ are assumed to have a function type $\alpha \to \beta$, and $u$ and $v$ are assumed to have type $\alpha$.\strut
\begin{center}
  \AXM{\strut s = t}
  \UIM{\strut (\Lam x s) = (\Lam x t)}
  \DP
  \quad
  \AXM{\strut s = t}
  \AXM{\strut u = v}
  \BIM{\strut s \, u = t \, v}
  \DP
\end{center}
For $\beta$-reduction, we add axioms $(\Lam x t) \, s = t [s / x]$. It is not hard to show that an equation $s = t$ is provable in this system if and only if $s$ and $t$ have the same $\beta$-normal form. If we add axioms $(\Lam x {t \, x}) = t$, it is similarly not hard to show that $s = t$ is provable if and only if $s$ and $t$ have the same $\beta\eta$-normal form.

Using the first $\eta$ axiom, the following is a derived rule, where the variable $x$ is not free in $s$ or $t$:\strut
\begin{center}
\AXM{\strut s \, x = t \, x}
\UIM{\strut s = t}
\DP
\end{center}
To see this, notice that from the premise we can derive $s = (\Lam x {s \, x}) = (\Lam x {t \, x}) = t$. This rule can be viewed as expressing a principle of \emph{extensionality}, since it says that if $s$ and $t$ are functions that have the same values for every input, then $s$ and $t$ are equal. As a result, the equational theory corresponding to $\beta\eta$-reduction is called the \emph{extensional} theory of the simply typed lambda calculus, in contrast to the \emph{intensional} theory corresponding to $\beta$-reduction alone.

\subsection{Extensions}
\label{subsection:lf:simply:typed:lambda:calculus:extensions}

The simply typed lambda calculus can be extended in various ways. For example:
\begin{itemize}
\item We can add a basic type $\ty{Nat}$ of natural numbers with constants $0 : \ty{Nat}$, $\fn{succ} : \ty{Nat} \to \ty{Nat}$ and \emph{recursors} $\fn{R}$ of appropriate types such that $\fn{R} \, f \, g \, 0$ reduces to $f$ and $\fn{R} \, f \, g \, (\fn{succ} \, x)$ reduces to $g \, x \, (\fn{R} \, f \, g \, x)$.
\item We can add a basic type $\ty{Bool}$ of booleans with constants $\fn{tt} : \ty{Bool}$ and $\fn{ff} : \ty{Bool}$ and a conditional function $\fn{cond}$ such that $\fn{cond} \, f \, g \, \fn{tt}$ reduces to $f$ and $\fn{cond} \, f \, g \, \fn{ff}$ reduces to $g$.
\item We can add \emph{product types} $\alpha \times \beta$ whose elements are ordered pairs $(s, t)$ with $s : \alpha$ and $t : \beta$. In addition to the pairing operation, product types come equipped with projections $\pi_0$ and $\pi_1$ such that $\pi_0 \, (s, t)$ reduces to $s$ and $\pi_1 \, (s, t)$ reduces to $t$. One sometimes also adds an analogue of $\eta$ reduction whereby $(\pi_0 \, p, \pi_1 \, p)$ reduces to $p$ for any $p : \alpha \times \beta$.
\item We can add \emph{sum types} $\alpha + \beta$, whose elements represent either an element of $\alpha$ or an element of $\beta$, tagged to indicate which is the case. These types come equipped with functions $\iota_0 : \alpha \to \alpha + \beta$ and $\iota_1 : \beta \to \alpha + \beta$ that insert elements from $\alpha$ and $\beta$, respectively. Each sum type also come with a $\fn{cases}$ function: given $f : \alpha \to \gamma$ and $g : \beta \to \gamma$, $\fn{cases} \, f \, g \, (\iota_0 \, a)$ reduces to $f \, a$ and $\fn{cases} \, f \, g \, (\iota_1 \, b)$ reduces to $g \, b$.
\end{itemize}
All these extensions preserve strong normalization and confluence. Notice the common features: in each case, we add a new type construction, canonical means of defining new elements of the new types, and a recursion principle with cases for each canonical construction.

With these extensions, the simply typed lambda calculus begins to look more like a programming language, one in which every program is guaranteed to terminate. Simple type theory, discussed in Section~\ref{section:lf:simple:type:theory}, adds a type $\Prop$ of propositions (or truth values) and can be seen as an even more far-reaching extension of the simply typed lambda calculus.

\subsection{Semantics}
\label{subsection:lf:simply:typed:lambda:calculus:semantics}

There are various kinds of semantics for the simply typed lambda calculus that are similar to the model-theoretic semantics for first-order logic. An \emph{extensional model} interprets each basic type $\alpha$ as a set $\tval{\alpha}$, and each type $\alpha \to \beta$ as a set of functions from $\tval{\alpha}$ to $\tval{\beta}$, in such a way that there are enough functions to interpret the lambda terms. In the \emph{full set-theoretic model}, $\tval{\alpha \to \beta}$ is interpreted as the set of \emph{all} functions from $\tval{\alpha}$ to $\tval{\beta}$, but there are smaller models, for example, where function types are interpreted as suitable sets of computable functions. The simply typed lambda calculus can also be interpreted in any \emph{cartesian closed category} in such a way that types are interpreted as objects and terms are interpreted as arrows. The extensional equational theory described in Section~\ref{subsection:lf:simply:typed:lambda:calculus:equational:theories} is sound and complete for all of these.

It is harder to describe a complete semantics for the intensional theory, since it is possible that $(\Lam x {s \, x}) = (\Lam x {t \, x})$ is derivable for a variable $x$ that does not occur in $s$ or $t$ but $s = t$ is not. \emph{Domain theory} provides one way of going about it.

In the theory of programming languages, approaches like these, which assign a mathematical object to every expression, are instances of \emph{denotational semantics}. In contrast, an \emph{operational semantics} gives meaning to a programming language by describing its computational behavior. For example, one can take the meaning of the simply typed lambda calculus to be given by a particular evaluation strategy for terms. Alternatively, one can take the meaning to be given by the family of evaluation strategies corresponding to $\beta$- or $\beta\eta$-reduction. In this sense, the meaning of a term is a specification of how we compute with it.

The \emph{Curry--Howard correspondence}, or \emph{propositions-as-types interpretation}, can be used to translate a computational interpretation of a system like the simply typed lambda calculus to a computational interpretation of a corresponding system of logical deduction. According to the informal \emph{Brouwer--Heyting--Kolmogorov} interpretation of intuitionistic logic, a proof of $A \Implies B$ is a procedure that maps any proof of $A$ to a proof of $B$, a proof of $A \And B$ is a pair consisting of a proof of $A$ and a proof of $B$, a proof of $A \Or B$ is either a proof of $A$ or a proof of $B$ tagged to indicate which is the case, and so on. The Curry--Howard isomorphism makes this precise by associating with each proposition in a logical system a type in a corresponding system of types, in such a way that proofs in the logical system correspond to terms in a calculus with those types.

To illustrate, let us consider an alternative presentation of the simply typed lambda calculus with products, in which types are assigned to variables by a finite set or sequence of specifications $x : \alpha{,}\; y : \beta{,}\; \ldots$ known as a \emph{context}. If $\Gamma$ is a context, a \emph{typing judgment} $\Gamma \proves t : \alpha$ says that $t$ has type $\alpha$ assuming the variables occurring in $t$ have the types specified by $\Gamma$. With these conventions, the formation rules for terms look like this:\strut
\begin{center}
\AXN{\strut\Gamma{,}\; x : \alpha \fCenter t : \beta}
\UIN{\strut\Gamma \fCenter \Lam x t : \alpha \to \beta}
\DP
\quad\quad
\AXN{\strut\Gamma \fCenter t : \alpha \to \beta}
\AXN{\strut\Gamma \fCenter s : \alpha}
\BIN{\strut\Gamma \fCenter t \, s : \beta}
\DP \\
\medskip
\AXN{\strut\Gamma \fCenter s : \alpha}
\AXN{\strut\Gamma \fCenter t : \beta}
\BIN{\strut\Gamma \fCenter (s, t) : \alpha \times \beta}
\DP
\quad\quad
\AXN{\strut\Gamma \fCenter t : \alpha \times \beta}
\UIN{\strut\Gamma \fCenter \pi_0 \, t : \alpha}
\DP
\quad\quad
\AXN{\strut\Gamma \fCenter t : \alpha \times \beta}
\UIN{\strut\Gamma \fCenter \pi_1 \, t : \beta}
\DP
\end{center}
Now notice that if we replace the types by formulas, function types by implications, and cartesian products by conjunctions, these are exactly the rules in natural deduction for implication and conjunction with the addition of the terms in the conclusion. If we read the term $t : A$ as saying that $t$ is a proof of $A$, we can interpret the simply typed lambda calculus as a calculus of \emph{proof terms} for the corresponding fragment of intuitionistic logic.

Thus, when propositions correspond to types, proofs correspond to terms. A procedure for normalizing terms in the simply typed lambda calculus becomes a procedure for normalizing proofs, and we can therefore take the meaning of an intuitionistic proof to be given by its computational behavior.

The correspondence can be extended to other logical constructions. For example, disjunction corresponds to sum types. To extend the correspondence to first-order logic, an equation $s = t$ should correspond to a type, $\ty{Id} \, s \, t$. But now notice that these identity types are parameterized by terms, which is to say, they are \emph{dependent types}. A universally quantified proposition $\Forall x A$ corresponds to a \emph{Pi type}, $\PiT x \alpha$, where $\alpha$ can depend on $x$. Similarly, $\Exists x A$ corresponds to a \emph{Sigma type}, $\SigmaT x \alpha$. We will return to the syntax of dependent type theory in Section~\ref{section:lf:dependently:typed:foundations}.

A \emph{realizability relation} provides another way of associating computational content with derivations. The general idea is to define a relation ``\emph{$e$ realizes $A$}'' between formulas $A$ and suitable data $e$, and show that whenever $A$ is provable in a certain deductive system, there exists an $e$ that realizes it. In Stephen Kleene's original version of realizability, $A$ is a formula in the language of first-order arithmetic and $e$ is a natural number. A realizer for $A \Implies B$ is an index for a computable function that takes any number realizing $A$ to a number realizing $B$, a realizer for $A \And B$ is a number coding a pair consisting of a realizer for $A$ and a realizer for $B$, and so on.

The propositions-as-types correspondence can be viewed as a kind of realizability in which the formula $A$ itself is a specification of the type of realizers, the typing judgment $e : A$ says that $e$ realizes $A$, and the system used to prove $A$, the system used to define $e$, and the system used to establish that $e$ realizes $A$ are all one and the same. Approaches based on realizability are more flexible in that they allow us to tease these components apart.

\section{Set Theory}
\label{section:lf:set:theory}

Set theory was designed in the early twentieth century to provide a uniform foundation for the methods of modern mathematics, including forms of algebraic abstraction and nonconstructive reasoning about infinite domains that were introduced in the late nineteenth century. It has the advantage that the underlying logic, first-order logic, is simple, and the axioms are intuitive and plausible. It has been so successful at describing modern methods that many mathematicians think of it as providing the gold standard for mathematical proof.

\subsection{The axioms}
\label{subsection:lf:set:theory:the:axioms}

The axioms of \emph{Zermelo--Fraenkel set theory} are expressed in classical first-order logic in a language with a single binary relation symbol, $\in$. We think of the entire mathematical universe as consisting of nothing but sets, and if $x$ and $y$ are sets, $x \in y$ says that $x$ \emph{is an element of} $y$.

The first axiom says that two sets are equal if they have the same elements:
\begin{itemize}
\item Extensionality:\enskip $\Forall{x, y} {(\Forall{z}{z \in x \Iff z \in y}) \Implies x = y}$.
\end{itemize}
Alternatively, we can formulate set theory in first-order logic without equality and take the antecedent of the implication as a definition of $x = y$. In that case, we need the following axiom in order to prove the substitution property of equality:
\begin{itemize}
\item $\Forall{x, y} {x = y \Implies \Forall{z} {x \in z \Iff y \in z}}$.
\end{itemize}

The next four axioms postulate the existence of the empty set, denoted $\emptyset$, with no elements; the unordered pair $\{ x, y \}$, whose elements are $x$ and $y$; the union $\bigcup x$, whose elements $z$ are exactly the members of some set $w$ in $x$; and the power set, $\mathcal{P}(x)$, whose members are exactly the subsets of $x$. In the statement of the power set axiom, $z \subseteq x$ abbreviates $\Forall {w \in z} {w \in x}$.

\begin{itemize}
\item Empty set:\enskip $\Exists x {\Forall y {y \notin x}}$.

\item Pairing:\enskip $\Forall {x, y} {\Exists z {\Forall w {w \in z \Iff w = x \Or w = y}}}$.

\item Union:\enskip $\Forall x {\Exists y {\Forall z {z \in y \Iff \Exists w {w \in x \And z \in w}}}}$.

\item Power set:\enskip $\Forall x {\Exists y {\Forall z {z \in y \Iff z \subseteq x}}}$.
\end{itemize}

The next axiom is a schema. For each formula $A(w, \vec x)$, it says that for any choice of parameters $\vec x$ and any set $y$, there is a set $\{ w \in y \mid A(w, \vec x) \}$ whose elements are exactly the elements $w$ of $y$ that satisfy $A$.
\begin{itemize}
\item Separation:\enskip $\Forall {\vec x, y} {\Exists z {\Forall w {w \in z \Iff w \in y \And A(w, \vec x)}}}$.
\end{itemize}
The axiom is restricted to separating elements from another set $y$ to avoid Russell's paradox: an unrestricted comprehension axiom would allow us to build the set $\{ w \mid w \not\in w \}$. The variable $z$ is not allowed to appear in $A$; otherwise, taking $A(w)$ to be $w \not\in z$ and taking $y$ to be any nonempty set would lead to a contradiction.

The next axiom asserts the existence of an infinite set:
\begin{itemize}
\item Infinity:\enskip $\Exists x \emptyset \in x \And \Forall y y \in x \Implies y \cup \{ y \} \in x$.
\end{itemize}
We will discuss this axiom further in Section~\ref{subsection:lf:inductively:defined:sets} and explain how it allows us to define the natural numbers.

Nothing so far rules out the possibility that a set can be an element of itself. The axiom of foundation does so. In words, it says that every nonempty set $x$ has an $\in$-least element $y$, that is, an element $y$ such that there is no $z$ in $x$ such that $z$ is in $y$.
\begin{itemize}
\item Foundation:\enskip $\Forall x {(\Exists y {y \in x}) \Implies \Exists {y \in x} {\Forall {z \in x} {z \notin y}}}$.
\end{itemize}
The axiom of foundation does not add logical strength. In set theory without that axiom, we can define what it means for a set to be \emph{well founded} and show that the well-founded sets again satisfy the axioms of set theory, as well as the axiom of foundation.

Once we have defined the set $\bN$ of natural numbers, we can use the power set axiom repeatedly to define sets $\mathcal{P}(\bN)$, $\mathcal{P}(\mathcal{P}(\bN))$, $\mathcal{P}(\mathcal{P}(\mathcal{P}(\bN)))$, \ldots. But it is consistent with the axioms we have seen so far that there is no single set that contains all of these. The next axiom schema implies that such a set exists.
\begin{itemize}
\item Replacement:
\[
  \Forall {x, \vec y} {(\Forall {z \in x} \ExistsUnique {w} {A(z, w, \vec y)) \Implies \Exists u {\Forall w {w \in u \Iff \Exists {z \in x} {A(z, w, \vec y)}}}}}.
\]
\end{itemize}
Here, $A$ is any formula that does not contain $u$. The schema says that if $A$ describes a function on $x$, then the image of $x$ under that function is again a set.

This constitutes the core axioms of Zermelo--Fraenkel set theory, sometimes abbreviated $\na{ZF}$. \emph{Zermelo--Fraenkel set theory with choice}, abbreviated $\na{ZFC}$, adds the following principle:
\begin{itemize}
\item Choice:\enskip $\Forall x {\emptyset \notin x \Implies \Exists {f \in (x \to \bigcup x)} {\Forall {y \in x} f(y) \in y}}$.
\end{itemize}
In this statement $x \to \bigcup x$ denotes the set of functions from $x$ to $\bigcup x$, relying on the representation of a function as a set of ordered pairs, as described in the next subsection. In words, it says that if $x$ is a collection of nonempty sets, there is a function $f$ which selects an element $f(y) \in y$ from each element $y$ of $x$.

The use of the notation $\mathcal P(x)$ for power sets and $\bigcup x$ for unions can be be viewed as definitional extensions, as described in Section~\ref{subsection:lf:definitional:extensions}. But now we have to contend with the slightly confusing situation that there are function symbols and relation symbols in the metatheory, which correspond to functions and relations on the universe of any model, as well as functions and relations as set-theoretic objects \emph{within} the theory, which are elements of the universe of any such model. It might help to think of the former as \emph{global} functions and relations associated with their defining formulas.

In particular, if $f$ is a function within the theory and $x$ is a set, the notation $f(x)$ refers to the unique element $y$ such that $(x, y)$ is in $f$. It should therefore be viewed as shorthand for a term of the form $\fn{app}(f, x)$, where $\fn{app}$ is a metatheoretic function symbol introduced to describe that value. The concerns about partial functions raised in Section~\ref{subsection:lf:partial:functions:and:undefined:terms} come into play; to introduce $\fn{app}(f, x)$, we must ensure that its defining formula describes a total function on the entire universe. We therefore have to choose a default value, such as the empty set, for elements $x$ that are not in the domain of $f$.

\subsection{Working in set theory}
\label{subsection:lf:working:in:set:theory}

The axioms of set theory provide recipes for constructing sets. Following Kazimierz Kuratowski we can define the \emph{ordered pair} $(x, y)$ to be the set $\{ \{ x \}, \{ x, y \} \}$, and then we can prove that $(x, y) = (x', y')$ if and only if $x = x'$ and $y = y'$. The set $A \times B$ is the set of all ordered pairs consisting of an element of $A$ and an element of $B$. Triples, quadruples, and so on can be defined by iterating these. A \emph{binary relation} $R$ on $A$ and $B$ is a subset of $A \times B$, and similarly for ternary relations and so on. A \emph{function} $f$ from~$A$ to~$B$ is a binary relation $R_f$ on $A$ and $B$ such that for every $x$ in $A$ there is a unique $y$ in $B$ such that $R(x, y)$ holds, and for any such $x$, $f(x)$ denotes that unique $y$. The set of all functions from $A$ to $B$ is denoted $A \to B$. A $k$-place function $f$ from $A_0, \ldots, A_{k-1}$ to $B$ can be interpreted as an element of $A_0 \times \ldots \times A_{k-1} \to B$.

If $f$ is a function from $A$ to $B$, the set $A$ is known as the \emph{domain} of $f$ and $B$ is known as the \emph{codomain}. The domain is a bona fide property of the function: it is the set of values $x$ such that for some $y$, $(x, y) \in f$. But the intended codomain has to be specified independently, since, for example, the function $f(x) = x^2$ defined on the real numbers, $\bR$, can be viewed as a function from $\bR$ to $\bR$ or as a function from $\bR$ to $\bR^{\ge 0}$, the nonnegative reals. The set of values $y$ such that $(x, y) \in f$ for some $x$ is called the \emph{image} of $f$. The term \emph{range} is sometimes used to refer to the image, and sometimes the codomain.

Dropping the existence requirement in the definition of a function gives rise to the notion of a \emph{partial function} from $A$ to $B$; in other words, a partial function $f$ from $A$ to $B$ is a relation $R_f$ such that for every $x$ in $A$ there is at most one $y$ in $B$ such that $(x, y)$ in $f$. Equivalently, a partial function from $A$ to $B$ is a function from some subset of $A$ to $B$.

The \emph{quotient construction} is central to modern mathematics, and an important tool in the use of mathematical abstraction. Let $A$ be any set, and let $\equiv$ be an \emph{equivalence relation} on $A$, that is, a reflexive, symmetric, transitive relation on $A$. Think of an equivalence relation in $A$ as a notion of what it means for two objects in $A$ to be the same, up to representational details that are irrelevant to a problem at hand. For any object $a \in A$, its \emph{equivalence class} $[a]$ is the set $\{ b \mid b \equiv a \}$, and the set $A / \equiv$ is the set of equivalence classes of elements of $A$. A function $f$ from $A$ to another set~$B$ is said to \emph{respect} the equivalence relation if whenever $x \equiv x'$, $f(x) = f(x')$. Such a function gives rise to another function $\bar f : {A / \equiv} \to B$ defined by $\bar f([a]) = f (a)$.

In Section~\ref{subsection:lf:inductively:defined:sets}, we will explain how to define the natural numbers. The integers, $\bZ$, the rationals, $\bQ$, and the real numbers, $\bR$, can then all be defined as suitable quotients. For example, the real numbers can be constructed as a quotient of the set of Cauchy sequences of rationals.

Finally, algebraic structures can be constructed as tuples. For example, a group is a tuple $(G, \circ, e, \cdot^{-1})$ satisfying the group axioms, a metric space $(M, d)$ consists of a set $M$ and a function $d : M \times M \to \bR$ satisfying the axioms for a metric space, and a topological space $(X, \mathcal{O})$ consists of a set $X$ together with its \emph{topology}, that is, a collection $\mathcal{O}$ of \emph{open} subsets of $X$ closed under finite intersections and arbitrary unions.

\subsection{Inductively defined sets}
\label{subsection:lf:inductively:defined:sets}

The natural numbers can be characterized abstractly as a set, $\bN$, with a constant, $0$, and a function $\fn{succ} : \bN \to \bN$ satisfying the following:
\begin{itemize}
  \item $0$ is not equal to $\fn{succ}(x)$ for any $x$ in $\bN$.
  \item The function $\fn{succ}(x)$ is injective.
  \item If $y \subseteq \bN$ contains $0$ and is closed under $\fn{succ}(x)$, then $y = \bN$.
\end{itemize}
The third principle is the principle of induction. It is not hard to show that these properties characterize the natural numbers in the sense that if $\bN'$, $0'$, and $\fn{succ}'(x)$ satisfy the same description, there is a bijection $F : \bN \to \bN'$ such that $F(0) = 0'$ and for every $x$, $F(\fn{succ}(x)) = \fn{succ}'(F(x))$. So we only need to show that the axioms of set theory imply the existence of one such set.

The von Neumann set-theoretic construction of the natural numbers runs as follows. Take $0$ to be the empty set, $\emptyset$, and take $\fn{succ}(x)$ to be $x \cup \{ x \}$. Notice now that the axiom of infinity says that there exists a set $y$ that contains $0$ and is closed under $\fn{succ}$. Call such a set \emph{inductive}. To satisfy the axiom of induction, roughly, we want to define $\bN$ to be the \emph{smallest} inductive subset of $y$. We can achieve that by defining it to be the intersection of all of them:
\[
  \bN = \bigcap \{ z \subseteq y \mid \mbox{$z$ is inductive} \}
\]
It is now an exercise in unraveling definitions to show that $\bN$ is inductive. To see that it satisfies the axiom of induction, let $z \subseteq \bN$ be inductive. Since $\bN$ is the intersection of all inductive subsets, it is a subset of $z$, and hence $\bN = z$.

It is clear from these definitions that $0$ is not equal to $\fn{succ}(x)$ for any $x$, since the empty set does not have any elements and $\fn{succ}(x)$ has at least one. It takes a bit more effort to show that $\fn{succ}(x)$ is injective on $\bN$, but it can be done. We therefore have a suitable definition of the natural numbers.

This is a classic example of an \emph{impredicative} construction: the set of natural numbers is defined with respect to a collection of sets---the collection of inductive subsets of $y$---that includes the very set, $\bN$, that is being defined. In the wake of the set-theoretic paradoxes around the turn of the twentieth century, this circularity raised concerns. The definition is problematic if you think of sets as somehow being created by their definitions. The problem goes away if you think of the definition as merely picking out a particular set from among a well-defined totality, namely, the collection of all subsets of $y$.

The principle of induction can be used to justify the principle of recursive definition: given any set $B$, any element $b \in B$, and any function $g : \bN \times B \to B$, there is a unique function $f$ satisfying
\begin{align*}
f(0)  &= b \\
f(x+1) &= g(x, f(x)) \quad \mbox{for every $x$ in $\bN$}.
\end{align*}

The structure $(\bN, \fn{succ}, 0)$ is an instance of an \emph{freely generated inductively defined structure}. The fact that it is inductively defined means that that the elements are generated from the bottom up by the constructors, in this case, $\fn{succ}(x)$ and $0$. This is sometimes thought of as saying that the structure contains \emph{no junk}, in the sense that there is nothing there except what \emph{has} be there to be closed under $0$ and $\fn{succ}$. The fact that $\fn{succ}$ is injective and $0$ is not in its range says that the structure contains \emph{no noise}, in the sense that no unexpected equations hold between elements. This property is essential to justifying the principle of recursive definition.


We will see in Chapter~4 that inductively defined structures are fundamental to interactive theorem proving. In some foundational frameworks, such as the systems of dependent type theory discussed in Section~\ref{section:lf:dependently:typed:foundations}, they are given as axiomatic primitives. In others, including most implementations of simple type theory, they are constructed. In that case, the methods generally follows the pattern described here: one finds a set or domain of objects big enough to support the construction, and then carves out the inductive structure with an impredicative definition.

\subsection{Variations}
\label{subsection:lf:set:theory:variations}

An axiom of first-order logic like $A \And B \Implies A$ is schematic, in that $A$ and $B$ can be replaced by arbitrary formulas. The separation and replacement axioms of $\na{ZFC}$ are schemas as well. They are best formulated, therefore, in a system that allows schematic presentations of axioms, with variables that can be replaced by suitable formulas. Some implementations of set theory, like those of Metamath and Isabelle/ZF, allow one to prove schematic theorems as well.

Other presentations of set theory avoid the need for schemata by positing two types of objects, sets and \emph{classes}. One can think of a class as a collection of sets that is potentially too large to be a set itself; in particular, the collection of all sets satisfying a first-order formula of $\na{ZFC}$ should form a class. There are two types of extensions of this sort. The theory $\na{NBG}$, or \emph{von Neumann--Bernays--G\"odel set theory}, is a conservative extension of $\na{ZFC}$, which means that it proves the same statements in the smaller language. In contrast, a theory like \emph{Morse--Kelley} set theory is properly stronger than $\na{ZFC}$, which is to say, the use of classes provides new consequences in the original language.

Another way to increase the strength of set theory is to postulate the existence of \emph{universes} of sets. A set $U$ is a universe if the following hold:
\begin{itemize}
  \item For every $x$ in $U$, $\mathcal{P}(x)$ is in $U$.
  \item For every $x$ in $U$ and $y \subseteq x$, $y$ is in $U$.
  \item If $x \subseteq U$ has cardinality less than that of $U$, then $x$ is in $U$.
\end{itemize}
In other words, a universe is closed under adding the power set of any element, adding all the subsets of an element, and adding all the subsets of the universe of smaller cardinality. \emph{Tarski--Grothendieck set theory} is $\na{ZFC}$ together with an axiom that says that every set $x$ is an element of some universe $U$.

The axiomatizations of set theory in both \relax{Mizar} and \relax{Metamath} include such universes, and this provides a way of working around the fact that some classes of objects are too large to be sets. For example, if we define a category to consist of a set of objects and a set of morphisms between them, we cannot talk about the category of groups, because the class of groups is too big to be a set. But we \emph{can} talk about the category of groups in any particular universe, and the closure properties on universes make that just as useful in practice.

Constructive theories based on the notion of set have also been studied \cite{MR519801,MR786465}. If we replace the axiom of infinity by its negation, the result is a theory of the \emph{hereditarily finite sets}, which is equi-interpretable with classical first-order arithmetic.

\section{Simple Type Theory}
\label{section:lf:simple:type:theory}

There are at least three ways to think of simple type theory. One way is to think of it as a variant of set theory that assigns a type to every object and separates the construction of objects from the construction of types. Mathematicians generally distinguish between domains of objects, like the set of natural numbers, function spaces, and algebraic structures, and elements of those domains, like the number 7, the exponential function on the reals, and the identity element of a group. Just as set theory provides fundamental recipes for constructing sets, type theory provides recipes for constructing such domains and their elements.

A second way to think of simple type theory is as an extension of first-order logic that allows us to quantify over relations and functions. \emph{Second-order logic} allows for quantifying over relations and functions on the first-order universe, and \emph{higher-order logic} allows for quantifying over relations and functions defined on these second-order domains, functions and relations on those third-order domains, and so on.

Yet a third way to think of simple type theory is as an extension of the simply typed lambda calculus with a new type, $\Prop$, intended to denote a type of propositions. Since a relation can be viewed as a function that returns a proposition, this provides a flexible language for describing functions and relations and making assertions about them.

All three perspectives come together in the formal systems we present in Section~\ref{subsection:lf:formulations:of:simple:type:theory}. We will start by exploring the second perspective with a discussion of second- and higher-order logic.

\subsection{Second-order logic}
\label{subsection:lf:second:order:logic}

Second-order logic can be viewed as a many-sorted logic in which we introduce, for each $n$, a sort ranging over $n$-place relations on a first-order universe and a function $\fn{app}_n(R, \vec x)$ to denote the result $R(\vec x)$ of applying the relation $R$ to elements $\vec x$ in that universe. We can then quantify over predicates and relations. For example, suppose we adopt a language for talking about undirected graphs, with a relation $E(x, y)$ that is intended to say that there is an edge from $x$ to $y$. Then the following second-order formula says that the graph is disconnected:
\[
\Exists {x, y, P} {P(x) \And \neg P(y) \And \Forall {u, v} {E(u, v) \Implies (P(u) \Iff P(v))}}
\]
\emph{Second-order logic} is many-sorted first-order logic in such a language, together with a \emph{comprehension} schema that says that every formula defines a relation:
\[
\Exists {R} {\Forall {\vec x} {R(\vec x) \Iff A(\vec x)}}
\]
In this chapter, when we present an axiom schema like this, it is to be understood that a formula like $A$ may have free variables other than the ones shown, in which case these variables are implicitly universally quantified. We require that $R$ is not a free variable of $A$. Without that stipulation, we would have the existence of a relation satisfying $R(\vec x) \Iff \neg R(\vec x)$, a contradiction.

Given any formula $A$ that contains an $n$-ary relation variable $R$ and given any formula $B(\vec x)$, we can imagine replacing every atomic formula $R(t_0, \ldots, t_{n-1})$ occurring in $A$ by the formula $B(t_0, \ldots, t_{n-1})$. It is convenient to use $A[\Lam {\vec x} {B} / R]$ to denote this replacement, even though the syntax of second-order logic, as we have just described it, does not include such lambda expressions. In a natural deduction formulation, the following rules can be derived from the comprehension axiom for the formula $B$:\strut
\begin{center}
\AXN{\strut\Gamma \fCenter A[\Lam {\vec x} {B} / R]}
\UIN{\strut\Gamma \fCenter \Exists R A}
\DP
\quad
\AXN{\strut\Gamma \fCenter \Forall R A}
\UIN{\strut\Gamma \fCenter A[\Lam {\vec x} {B} / R]}
\DP
\end{center}
Conversely, one can derive the comprehension schema from the first of these rules.

If we include an equality symbol for each of the new sorts, we may want to add the corresponding axioms of \emph{extensionality}:
\[
\Forall {R, R'} {(\Forall {\vec x} {R(\vec x) \Iff R'(\vec x)}) \Implies R = R'}
\]
With the substitution axiom for equality, this means that whenever two relations agree on all possible arguments, each can be substituted for the other. In the context of second-order logic, we can instead \emph{define} $R = R'$ by the formula $\Forall {\vec x} {R(\vec x) \Iff R'(\vec x)}$. With this definition, the usual equality axioms are derivable.

In second-order logic, we can prove $\Forall {x, y} x = y \Iff \Forall P {P(x) \Iff P(y)}$. This says that two elements of the first-order universe are equal if and only if they have all the same properties, a characterization known as \emph{Leibniz equality} or \emph{identity of indiscernibles}. Alternatively, we can take the right-hand side of the equivalence as the definition of equality on the first-order universe, in which case the usual laws of equality are derivable.

Other simplifications are possible. We can think of $0$-ary relations as propositions, and comprehension gives us $\Forall {\vec z} {\Exists P {P \Iff A(\vec z)}}$, where $\vec z$ denotes the free variables of $A$. In other words, every formula corresponds to a proposition. We can then define all the logical connectives in terms of universal quantification and implication as follows:
\begin{align*}
A \And B & \; \equiv \; \Forall P {(A \Implies (B \Implies P)) \Implies P} \\
A \Or B  & \; \equiv \; \Forall P {(A \Implies P) \Implies ((B \Implies P) \Implies P)} \\
\bot & \; \equiv \; \Forall P P \\
(\Exists x A) & \; \equiv \; \Forall P {(\Forall x {A \Implies P}) \Implies P}
\end{align*}
As a result, one can formulate second-order logic with only the universal quantifier and implication.

Moreover, given the correspondence between formulas and propositions, there is no good reason to distinguish between the two. In other words, we can conflate the propositional variable $P$ and the formula $\fn{app}_0(P)$ that says that $P$ holds. As a result, we can take formulas to be terms of the sort of propositions, which we now call $\Prop$. Implication is then a binary operation that takes two elements of $\Prop$ and returns an element of $\Prop$, and universal quantification is a variable-binding construction that applies to terms of sort $\Prop$.

It is reasonable to want to quantify over functions in addition to relations. This can be reduced to quantification over relations by interpreting an $n$-ary function variable~$f$ as an $(n+1)$-ary relation symbol $R_f$ and relativizing to \emph{functional} relations, that is, relations satisfying $\Forall {\vec x} {\ExistsUnique y {R(\vec x, y)}}$. We can then use the method of definite descriptions, presented in Section~\ref{subsection:lf:definitional:extensions}, to interpret uses of the function symbol $f$. Under this interpretation, both classical and intuitionistic versions of second-order logic can prove the following comprehension schema for functions:
\[
  (\Forall {\vec x} {\ExistsUnique y {A(\vec x, y)}}) \Implies \\
  \Exists {f} {\Forall {\vec x} {A(\vec x, f(\vec x))}}
\]
The stronger schema obtained by dropping the uniqueness requirements in the hypothesis is known as the axiom of choice.

In classical second-order logic, it is possible to go the other way and interpret relations as functions. Suppose that, in addition to the first-order universe, we introduce another first-order sort $\ty{Bool}$ of truth values, with constants $\top$ and $\bot$ and an axiom $\top \ne \bot$. Then, for any formula $A(\vec x)$, classical logic proves $\Forall {\vec x} {\ExistsUnique y {y = \top \Iff A(\vec x)}}$. If $f(\vec x)$ meets this specification of a map to $\ty{Bool}$, then $A(\vec x)$ is equivalent to $f(\vec x) = \top$. As a result, we can interpret any relation $R(t_0, \ldots, t_{n-1})$ as an equation $\chi_R(t_0, \ldots, t_{n-1}) = \top$, where $\chi_R$ is the characteristic function of $R$.

In intuitionistic logic, however, a two-valued sort is not sufficient to represent truth values. We have already noted that comprehension gives us $\Forall {\vec z} {\Exists P {P \Iff A(\vec z)}}$, where $P$ is of the second-order sort $\Prop$, so $\Prop$ can be viewed as a sort of truth values. But any map $\vec z \mapsto P$ is a function from the sort of individuals to the sort of propositions, which takes us beyond second-order logic. It is, however, within the scope of \emph{higher-order logic}, which we turn to next.

\subsection{Higher-order logic}
\label{subsection:lf:higher:order:logic}

Having opened the door to quantifying over predicates and relations on a first-order universe, there is no reason to stop there. One way to describe higher-order logic is to start with a collection $\mathcal B$ of basic sorts and define a set of higher-order sorts inductively as follows:
\begin{itemize}
  \item Every basic sort is a sort.
  \item If $\alpha_0, \ldots, \alpha_{n-1}$ are sorts, so is $[\alpha_0, \ldots, \alpha_{n-1}]$.
\end{itemize}
Here, $[\alpha_0, \ldots, \alpha_{n-1}]$ is intended to denote the sort of relations $R(x_0, \ldots, x_{n-1})$ where each $x_i$ is of the sort $\alpha_i$. If $A$ is any formula and the variables $\vec x$ range over the sorts $\vec \alpha$, we have the comprehension axiom $\Exists {R} {\Forall {\vec x} {R(\vec x) \Iff A(\vec x)}}$, where $R$ has sort $[\vec \alpha]$. As in the case of second-order logic, the comprehension axioms can be formulated instead in terms of the $\exists$-introduction rule.

A convenient alternative, favored by proof assistants, is to formulate higher-order logic as an extension of the simply typed lambda calculus. To interpret relations as functions, we just need to add a basic type $\Prop$ of propositions. In a natural deduction formulation, in addition to reflexivity, symmetry, and transitivity of equality, we add rules to express that equality is preserved by lambda abstraction and application, as described in Section~\ref{subsection:lf:simply:typed:lambda:calculus:equational:theories}. If we take formulas to be terms of type $\Prop$, we no longer need to add comprehension axioms, since we now have a term $\Lam {\vec x} {A(\vec x)}$ to denote the relation $R(\vec x)$ defined by $A(\vec x)$. The axiom for $\beta$-reduction gives $(\Lam {\vec x} {A(\vec x)}) \, \vec t = A(\vec t)$. To avoid a multiplicity of term-binding operations, we can introduce constants $\fn{Forall}_\alpha : (\alpha \to \Prop) \to \Prop$ and $\fn{Exists}_\alpha : (\alpha \to \Prop) \to \Prop$ for each type $\alpha$, define $\Forall x A$ to be $\fn{Forall}_\alpha \; \Lam x A$, and define $\Exists x A$ to be $\fn{Exists}_\alpha \; \Lam x A$.

In contrast to the second-order case, here we cannot easily eliminate the need for extensionality axioms. For example, combining the axiom of extensionality for unary second-order predicates with the substitution rule for equality yields the following for any \emph{third-order} relation $\mathcal{R}$:
\[
  (\Forall x {P(x) \Iff Q(x)}) \Implies (\mathcal{R}(P) \Iff \mathcal{R}(Q))
\]
In the second-order setting, we could define $P = Q$ to be $\Forall x {P(x) \Iff Q(x)}$, in which case the extensionality axiom holds by definition. But when we take equality as a primitive in the higher-order setting, we have $P = Q \Implies \mathcal{R}(P) = \mathcal{R}(Q)$ as an instance of the usual substitution axiom for equality; so if instead we define $P = Q$ to be $\Forall x {P(x) \Iff Q(x)}$, we have to add implications like the one above as additional axioms. In other words, there are two interpretations of an equation $P = Q$ to contend with: Leibniz equality, which says that $P$ and $Q$ can be exchanged in any formula, and extensional equality, which says that $P$ and $Q$ hold of the same elements. The first always implies the second, and extensionality is equivalent to saying that the second implies the first.

When we formulate higher-order logic in terms of the simply typed lambda calculus, there are two types of extensionality axioms to consider:
\begin{itemize}
\item Propositional extensionality:\enskip $\Forall {P, Q : \Prop} {(P \Iff Q) \Implies P = Q}$.
\item Function extensionality:\enskip $\Forall {f, g : \alpha \to \beta} {(\Forall x {f \, x = g \, x}) \Implies f = g}$.
\end{itemize}
As noted in Section~\ref{section:lf:simply:typed:lambda:calculus}, in the presence of the other equality axioms, function extensionality is equivalent to having an axiom $\Forall f (\Lam x {f \, x}) = f$ for each function type.

It turns out that whether we formulate higher-order logic with relations or functions, it is possible to interpret higher-order logic with extensionality in higher-order logic without it, by relativizing quantifiers and variables to the \emph{hereditarily extensional} objects. This shows that, in a theoretical sense, the principle of extensionality does not add logical strength. But the extensional view of functions is fundamental to classical mathematics, and most implementations of classical higher-order logic include some formulation of extensionality. In constructive dependent type theory, the issues are more subtle, since extensionality bears on a system's computational interpretation. We will return to this issue in Section~\ref{section:lf:dependently:typed:foundations}.

Regarding semantics, if we formulate second- and higher-order logic as systems of many-sorted first-order logic with relations, having a first-order model is equivalent to having interpretations of each relation sort with enough relations to satisfy the comprehension axioms. Such a model is known as a \emph{Henkin model}, and soundness and completeness for Henkin semantics carries over from first-order logic. Similar considerations hold for the formulation of higher-order logic in the simply typed lambda calculus with extensionality. (As in indicated in Section~\ref{subsection:lf:simply:typed:lambda:calculus:semantics}, since lambda abstraction takes us outside the realm of first-order logic, it takes more work to describe a semantics in which extensionality can fail but equality is nonetheless preserved by lambda abstraction.).

In the \emph{standard} interpretation of second-order logic, second-order quantifiers range over \emph{all} relations on the first-order universe, and similarly for higher-order functions and relations in higher-order logic. Because this semantics admits a definition of the natural numbers (modulo an assumption that the first-order universe is infinite), validity in this semantics is far from decidable. As a result, there is no computable notion of proof that gives rise to a system that is sound and complete for the semantics.

Intuitionistic higher-order logic has interesting algebraic and category-theoretic interpretations. It can be seen as the \emph{internal logic} of a \emph{topos}, and versions of intuitionistic logic are sound and complete with respect to semantics in an elementary topos. The latter is a generalization of the notion of a topos of sheaves on a Grothendieck site, and that, in turn, is a far-reaching generalization of the notion of a topos of sheaves on a topological space, first developed by Grothendieck for its applications to algebraic geometry.

\subsection{Formulations of simple type theory}
\label{subsection:lf:formulations:of:simple:type:theory}

To sum up the conclusions from the previous subsection, we can build a system of higher-order logic using terms in the simply typed lambda calculus with a basic type of propositions, $\Prop$, and another basic type of individuals, $\ty{Ind}$, with constants $\fn{Imp} : \Prop \to \Prop \to \Prop$ and $\fn{Forall}_\alpha : (\alpha \to \fn{Prop}) \to \fn{Prop}$ for each type $\alpha$. We write $\Forall x A$ for $\fn{Forall} \, \Lam x A$, define the other propositional connectives as described in Section~\ref{subsection:lf:higher:order:logic},
and define $x =_\alpha y$ at each type $\alpha$ by $\forall P : \alpha \to \Prop, P \, x \Iff P \, y$. We can take the proof system to establish sequents of the form $\Gamma \fCenter A$, where the elements of $\Gamma$ as well as a $A$ are terms of type $\Prop$. The axioms and rules of the proof system are as follows:
\begin{itemize}
  \item the natural deduction rules for assumption, implication, and the universal quantifier;
  \item the equational rules for the simply typed lambda calculus: reflexivity, transitivity, a rule expressing that lambda abstraction respects equality, and axioms for $\beta$-reduction and $\eta$-reduction;
  \item propositional extensionality: $(P \Iff Q) \to P = Q$.
\end{itemize}
Church used $\omicron$ instead of $\Prop$ and $\iota$ instead of $\ty{Ind}$, and this notation is still common in the literature. In classical systems it is possible to show that $\Prop$ has only two values, and so, in such systems, it can be called $\ty{Bool}$ instead.

An alternative approach, adopted by John Harrison in the HOL Light proof assistant \cite{DBLP:conf/tphol/Harrison09a}, is to start with the simply typed lambda calculus with basic types $\Prop$ and $\ty{Ind}$ as before, but now take equality as basic and define the logical connectives in terms of that. Specifically, we define the following:
\begin{align*}
  \top & \; \equiv \; (\Lam {P : \Prop} P) = (\Lam {P : \Prop} P) \\
  (\Forall x P) & \; \equiv \; (\Lam x P) = (\Lam x \top) \\
  P \And Q & \; \equiv \; (\Forall {\mathcal R} {\mathcal R \, P \, Q = \mathcal R \, \top \, \top}) \\
  P \Implies Q & \; \equiv \; P \And Q = Q
\end{align*}
In the definition of $P \And Q$, the variable $\mathcal R$ has type $\Prop \to \Prop \to \Prop$. Equality on $\Prop$ serves as bi-implication, and the other logical connectives can be defined as described in Section~\ref{subsection:lf:second:order:logic}. The axioms and rules of the proof system are then as follows:
\begin{itemize}
  \item Equality axioms and rules: reflexivity, transitivity, rules expressing that application and lambda abstraction respect equality, and, finally, $\beta$-reduction and $\eta$-reduction.
  \item The following logical rules, where $P$ and $Q$ are assumed to have type $\Prop$:\strut
\begin{center}
\AXN{\strut P \fCenter P}
\DP
\quad
\AXN{\strut\Gamma \fCenter P = Q}
\AXN{\strut\Delta \fCenter P}
\BIN{\strut\Gamma{,}\; \Delta \fCenter Q}
\DP
\quad
\AXN{\strut\Gamma{,}\; P \fCenter Q}
\AXN{\strut\Delta, Q \fCenter P}
\BIN{\strut\Gamma{,}\; \Delta \fCenter P = Q}
\DP
\end{center}
\end{itemize}

The two formulations of higher-order logic just described are equivalent and give rise to an intuitionistic logic. The axioms and rules presented so far have a (classical) model in which $\ty{Ind}$ consists of a single element. To obtain a stronger system, we add an \emph{axiom of infinity}, asserting that $\ty{Ind}$ is Dedekind-infinite:
\[
\Exists {f : \ty{Ind} \to \ty{Ind}} {(\Forall {x, x'} {f \, x = f \, x' \Implies x = x'}) \And \Exists y {\Forall x {f \, x \ne y}}}.
\]
This says that there exists a function from $\ty{Ind}$ to $\ty{Ind}$ that is injective but not surjective. Finally, we add a choice operator $\varepsilon_\alpha : (\alpha \to \Prop) \to \alpha$ with the following axiom, in which $P$ has type $\alpha \to \Prop$ and $x$ has type $\alpha$:
\[
\Forall {P, x} {P \, x \Implies P \, (\varepsilon \, P)}.
\]
This axiom says that if there is anything that satisfies the predicate $P$, $\varepsilon \, P$ chooses one such value. This operation is often called the \emph{Hilbert epsilon function}, since it was used in the context of first-order arithmetic by David Hilbert and his students in the early 1920s.

A trick due to Radu Diaconescu \cite{MR373893} can be used to derive the law of the excluded middle, $\Forall P {P \Or \neg P}$, from extensionality and choice. For a sketch of the argument, let $P$ be an arbitrary element of $\Prop$, and define the following two predicates:
\begin{align*}
 & U \; \equiv \; \Lam x {(x = \top) \Or P}
&& V \; \equiv \; \Lam x {(x = \bot) \Or P}
\end{align*}
From extensionality for propositions and functions, we have $P \Implies U = V$, and hence $P \Implies \varepsilon \, U =\varepsilon \, V$. From the choice axiom, we have $(\varepsilon \, U = \top) \Or P$ and $(\varepsilon \, V = \bot) \Or P$. These two disjunctions give rise to four cases. In three of them, $P$ holds, and in the fourth we have $\varepsilon \, U \ne \varepsilon \, V$, and hence $\neg P$.

Those who prefer to avoid using the choice operator can replace it with a definite description operator. To define the natural numbers, one should then add constants to name the objects asserted to exist by the axiom of infinity, and for classical logic one has to add the law of the excluded middle as a separate axiom.

\subsection{Extensions}
\label{subsection:lf:simple:type:theory:extensions}

The previous subsections provide a clean theoretical presentation of simple type theory. We now consider some extensions that are commonly adopted by proof assistants for practical purposes. Even though they extend the theoretical core conservatively, they are so fundamental to the operation of the system that it makes sense to view them as part of the foundation rather than incidental features of the implementation.

First, proof assistants invariably allow users to introduce \emph{definitions}. In the context of simple type theory, this amounts to allowing the user to introduce, corresponding to any term $t$ of type $\alpha$, a new constant symbol $\fn{D}$ of the same type, with the axiom $\fn{D} = t$. With lambda abstraction, this allows for explicit definitions of functions and relations. Together with either a definite or an indefinite description operator, this allows for implicit definitions as well. Most proof assistants also provide mechanisms to support things like infix notation, but, in contrast, we view that as syntactic sugar on top of the core logic.

Second, given a set of hypotheses, $\Gamma$, a formula, $A$, some variables $x_0, \ldots, x_{n-1}$, and terms $t_0, \ldots, t_{n-1}$ of the same type, most systems include a substitution rule:\strut
\begin{center}
\AXN{\strut\Gamma \fCenter A}
\UIN{\strut\Gamma[\vec t / \vec x] \fCenter A[\vec t / \vec x]}
\DP
\end{center}
This makes it convenient to store theorems schematically and instantiate them. This is a conservative extension because all the axioms and rules of the calculus, as we have presented them, are already closed under substitution.

Third, most systems also allow variables $\alpha, \beta, \gamma, \ldots$ ranging over types and a substitution principle for them:\strut
\begin{center}
  \AXN{\strut\Gamma \fCenter A}
  \UIN{\strut\Gamma[\vec \tau / \vec \alpha] \fCenter A[\vec \tau / \vec \alpha]}
  \DP
\end{center}
Here, $\vec \tau$ are arbitrary type expressions, possibly involving other variables. One has to spell what it means to substitute a type expression for a type variable, but that is straightforward. As an example, from the theorem $(\Lam {x : \alpha} x) \, y = y$ for any variable $y$ of type $\alpha$, we can conclude $(\Lam {x : \ty{Ind} \to \beta} x) \, y = y$ for any variable $y$ of type $\ty{Ind} \to \beta$.

Fourth, we have included $(\Lam x t) \, s = t [s / x]$ among the axioms of simple type theory, but some systems, like Isabelle, identify a $\beta$-redex with its contraction when comparing and matching expressions. This makes the axioms for $\beta$-reduction unnecessary, since they are just instances of the reflexivity of equality. Such an identification is one way of incorporating a broader notion of \emph{definitional equality} into the logical foundation, something which is discussed in greater detail in Section~\ref{subsection:lf:dependent:type:theory:syntax}.

Finally, proof assistants based on simple type theory generally provide a mechanism for defining new types by abstracting a nonempty subset of a type. This makes it possible to define types for the natural numbers, booleans, integers, rationals, reals, and so on. Type constructions can also be parameterized by type variables, so that $\alpha \times \beta$ is an operation that returns the cartesian product, and $\ty{List} \; \alpha$ is the type of lists of elements of objects of type $\alpha$.

To understand the principle for defining new types, you should keep in mind that the axioms and rules of higher-order logic, as we have presented them, presuppose that every type is nonempty. Indeed, for every type $\alpha$, $\varepsilon_\alpha \, (\Lam x \top)$ is a closed term of type $\alpha$
and $\Exists {x : \alpha} {x = x}$ is a theorem for every type $\alpha$. (There are variations of higher-order logic that allow for the possibility of empty types by maintaining a context of variables that have been assumed to be elements of each type. We will see that this idea is implemented organically in the context of dependent type theory, where, in contrast, it is hard to avoid the possibility of empty types.)

The principle for defining new types is as follows. Suppose $\tau$ is a type expression, possibly involving some type variables $\vec \alpha$. Let $P$ be a closed term of type $\tau \to \Prop$. (By \emph{closed} we mean that there are no ordinary variables free in $P$, although \emph{types} of $P$ and its subexpressions may depend on the type variables $\vec \alpha$.) Suppose moreover we have have proof that $P$ is nonempty, that is, a proof of $\Exists x {P \, x}$. Then we can introduce a new type constructor, $\ty{D} \, \vec \alpha$, abstractly representing the elements of $\tau$ satisfying $P$. We express this by providing a function $\fn{abs} : \tau \to \ty{D}$ which maps any element $x$ of $\tau$ satisfying $P$ to its abstraction in $\fn{D}$, a function $\fn{repr}$, which maps any element of $\ty{D}$ to its representative in $P$, and axioms saying that these provide a bijective correspondence. Operations on $D$ can be defined from operations on $\tau$ via these functions.

Seeing that such definitions yield a conservative extension is not entirely straightforward, but it should seem plausible that one can translate proofs in the expanded system to proofs in the original by relativizing quantifiers to the defining predicates. Alternatively, one can use a model-theoretic argument and show that one can expand any (Henkin) model for the system without definitions to the system with them, by iteratively adding new sorts to accommodate these definitions.

\subsection{Working in simple type theory}
\label{subsection:lf:working:in:simple:type:theory}

Given a type $\alpha$, subsets of $\alpha$ can be represented as predicates $P : \alpha \to \Prop$. In other words, we can define new types $\ty{Set} \, \alpha$ as abstractions of $\alpha \to \Prop$, or simply take $x \in P$ to be notation for $P \, x$. We then obtain expected properties of sets, with the caveat that when we talk about a set, it is always, implicitly, a subset of some underlying type. With this identification of predicates with sets, our principle for defining new types allows us to declare a new type in bijective correspondence with any nonempty subset of an existing type.

Up to a point, the development of mathematics in simple type theory looks much like the development of mathematics in set theory. Fixing the injective function $f$ asserted to exist by the axiom of infinity and an element $c$ that is not in its range, we can define the type of natural numbers as an abstraction of the intersection of all subsets of $\fn{Ind}$ containing $c$ and closed under $f$. Cartesian products $\alpha \times \beta$ can be represented by the set of relations $R : \alpha \to \beta \to \Prop$ with the property that for some $a : \alpha$ and $b : \beta$, we have $\Forall {x, y} R \, x \, y \Iff x = a \And y = b$. In other words, we represent pairs as binary relations on $\alpha$ and $\beta$ that hold of exactly one pair of elements. Iterating cartesian products, we get arbitrary tuples of data, sometimes called \emph{records}.

Quotients can also be defined as in set theory: if $\alpha$ is a type and $R$ is an equivalence relation on $\alpha$, we can define a new type $\alpha / R$ represented by the set of all nonempty equivalence classes modulo $R$. The integers, rationals, real numbers, and so on can be defined in this way. Inductively defined types like $\fn{List} \, \alpha$ can also be defined as in set theory.

Structures are generally parameterized by types. For example, a group structure on a type $\alpha$ is an element of $(\alpha \to \alpha \to \alpha) \times \alpha \times (\alpha \to \alpha)$ consisting of the group operation, the identity, and the inverse function. An element of this type is a group if it satisfies a predicate that says that the multiplication is associative and the identity and inverse have the required properties. Proving a theorem about arbitrary groups then amounts to proving theorems about group structures satisfying the predicate, where $\alpha$ is left as a variable ranging over types. This allows us to prove theorems about arbitrary groups, rings, and fields, and then instantiate them to particular instances, like the integers, rationals, and reals.

\subsection{Sets versus types}
\label{subsection:lf:sets:versus:types}

In comparison to set theory, typed frameworks offer both advantages and disadvantages. An important advantage is that functions and predicates that have been defined in a library come with a specification of the expected types of their arguments, so the system can provide users with an informative error messages when these expectations are not met. Type disciplines also support overloading and more concise manners of expression; when stating a theorem like $\Forall {x y : \bZ} {(x + y)^2 = x^2 + 2 \cdot x \cdot y + y^2}$, a user does not have to specify that the addition symbol denotes addition on $\bZ$, since that can be inferred from the types of the variables. In a set-theoretic framework, one has to use different notation for the different interpretations of $+$ or devise conventions and mechanisms to disambiguate the meaning.

Another advantage is that a proof assistant can use typing information to draw conclusions that generally have to be justified explicitly in set theory. For example, using facts from the library to justify the theorem above may require knowing that a large number of auxiliary expressions, like $(x + y) \cdot x$ and $x^2$, are integers. In a typed framework, this can be inferred automatically when checking that the expression is well formed. In a set-theoretic framework, either users have to do more work or system designers have to provide ad hoc automation to fill in the necessary steps.

Some of the disadvantages of using a type discipline have already been mentioned in Section~\ref{section:lf:overview}. A typed foundation is generally more complicated than an untyped one, making it harder to implement the framework and harder to ensure correctness. And from a mathematical perspective, the distinction between objects and types is unnatural. In ordinary mathematics, the symbol $\bZ$ for the integers is sometimes used as a type, to delimit the range of significance of a quantifier or a variable, and sometimes used an an object, something that can be manipulated and passed as an argument to a function. As a result, a typing discipline sometimes feels like more of a hindrance than a help, since it often distinguishes between objects that we would like to consider the same. It is disappointing that we have to worry about the differences between the natural number $2$, the integer $2$, and the real number $2$ when formalizing a theorem in type theory. In an implementation of set theory such as that of Mizar, they are the all same object, since the natural numbers and integers are defined as subsets of the real numbers.

There are a few things that can be said in defense of types. The first is that one does not \emph{have} to use them. One can certainly implement the natural numbers and integers using predicates $\fn{is\text{-}nat}$ and $\fn{is\text{-}int}$ on the type of real numbers, in which case formal statements about natural numbers and integers look much like they do in a set-theoretic setting. One can even axiomatically introduce a type of sets closed under the usual set-theoretic constructions \cite{DBLP:conf/itp/HanD19,DBLP:conf/cade/Harrison06,DBLP:conf/ictac/Obua06,DBLP:journals/access/SunY20} and develop most of one's mathematics on that type. The reason that this is not generally done is that types are often useful. So the real issue is deciding when it is advantageous to use them and when it is not.

A second consideration is that many of the challenges that have to do with identifying mathematical objects in different domains come up in set-theoretic frameworks as well. It might be possible to set up definitions so that the integer 2 and the real number 2 coincide, but there are also 2s in the quaternions, the octonians, every field extension of the rationals, and the $p$-adic numbers for any prime $p$. Mathematicians generally identify the integers with their images in any ring of characteristic zero. They are more generally notorious for identifying structures that are \emph{canonically isomorphic}, where canonical isomorphism is like pornography: it is hard to define, but they know it when they see it. One often refers to ``the two-element group'' even though, of course, there are many of them; in measure theory and probability one conflates the difference between functions and equivalence classes of functions up to almost-everywhere equivalence; and so on. Making these practices formal causes headaches for the set theorist and type theorist alike.

Moreover, in situations like these, type theory offers means to help. Given an expression $e$ of type $\bZ$ where an expression of type $\bR$ is expected, an implementation of type theory can silently or overtly insert a \emph{coercion} or \emph{cast} that embeds $\bZ$ into $\bR$, as is done in many programming languages. Types also support a mechanism known as \emph{type class inference} that can be used to infer algebraic structure that is implicitly associated with a type, as well as to overload notation in a principled way. Coercions and class inference can be used in conjunction with dependent type theory as well as simple type theory, and they are discussed in detail in Chapter~7.

\subsection{Limitations of simple type theory}
\label{subsection:lf:limitations:of:simple:type:theory}

Once we decide that types are useful, the question is then how elaborate we want our type system to be. Many-sorted first-order logic is a first step toward dividing expressions into syntactic categories, and simple type theory is the next. We should not look to extend a type system without good reason, so this subsection is dedicated to considering some of the limitations of simple types.

In simple type theory, there is a sharp distinction between objects and types, and types themselves are not objects. A theorem with polymorphic type variables $\alpha, \beta, \gamma, \ldots$ is implicitly universally quantified over those variables, but one cannot quantify over them explicitly. There is no way to even state the fact that there exists a type with two elements, let alone prove it. Since quantifying over algebraic structures requires, in particular, quantifying over their carrier types, simple type theory has limited capacities for reasoning about them as well.

Similarly, in simple type theory, functions between types are modeled as polymorphic type constructors like $\ty{List} \; \alpha$, which are not objects either. So it is impossible to reason about higher-order operations on types and algebraic structures. One can define the abstract notion of a \emph{category} in simple type theory using types to represent the objects and morphisms, but then one cannot use them to reason about concrete categories, like the category of groups. Simple type theory poses similar difficulties for reasoning about abstractions in computer science. Functional programming languages like Haskell rely on \emph{monads}, which are functors on types equipped with operations satisfying certain laws. In simple type theory, we cannot even define the notion of a functor, let alone prove things about them.

Another drawback of giving types and type constructors second-class status is that it is becomes impossible to define type constructions that depend on \emph{parameters} that are objects. Although we can define the type of lists, $\ty{List} \; \alpha$, we cannot define the type $\ty{Vector} \, \alpha \, n$ of tuples of elements of $\alpha$ of length $n$, where $n$ is a parameter of type~$\bN$. In mathematics, algebraic structures are often constructed from data:
\begin{itemize}
  \item For every $n > 0$, the structure $\bZ / n \bZ$ of integers modulo $n$ is a ring, and a field if $n$ is prime.
  \item Multivariate analysis deals with space $\bR^n$ of vectors of real numbers of length $n$.
  \item Linear algebra deals with $m \times n$ matrices, and, for any $n > 0$, the $n \times n$ matrices with elements in a given ring again form a ring.
  \item For every ring, $R$, we have the ring $R[X]$ of polynomials with coefficients in $R$.
  \item For every group, $G$, we can consider its group of automorphisms, $\fn{Aut}(G)$.
  \item A \emph{sheaf} over a topological space $T$ assigns an algebraic structure to each open set of $T$ and a restriction map to every inclusion.
\end{itemize}
Such constructions cannot be modeled straightforwardly as type constructors in simple type theory. These limitations also bear on applications to computer science, since we cannot define a type of $n$-bit words with $n$ as a parameter, or a type of search trees depending on an order function.


In short, the problem is that simple type theory prevents us from thinking about types as full-fledged mathematical objects, and sometimes that is what we want to do. This reflects a fundamental tension between two ways of thinking about types. On the one hand, types are useful syntactic categorizations, enabling a system to determine quickly whether an expression is well formed, fill in information that is left implicit, and provide useful error messages. From that point of view, a type system should be straightforward, deterministic, and efficient. On the other hand, we want flexibility when it comes to specifying systems and spaces of objects. This pushes toward a more elaborate type system.

Simple type theory offers a compromise, with a simple but expressive typing discipline. If we are unhappy with this compromise, we have to make a choice. One option is to abandon the use of types and stick with set theory. Dependent type theory, which we will turn to next, goes to the opposite extreme: it brings the expressivity of the language of objects to the language of types, with all the advantages and disadvantages that come with it.

\section{Dependent Type Theory}
\label{section:lf:dependently:typed:foundations}

As with simple type theory, there are various ways to think about dependent type theory. We can view it as a system for constructing mathematical objects and domains of objects, one that overcomes the limitations of simple type theory described in Section~\ref{subsection:lf:limitations:of:simple:type:theory}. But there is another way of thinking of dependent type theory that highlights its computational nature. We have seen that simple type theory can be viewed as an extension of the simply typed lambda calculus, and that, thanks to strong normalization and confluence, the latter can be viewed as a programming language. It is a rather restricted programming language, and the construction of the natural numbers that was described in Section~\ref{section:lf:simple:type:theory} does not directly extend the computational interpretation. But, as described in Section~\ref{subsection:lf:simply:typed:lambda:calculus:extensions}, we can augment the simple types with the types for the natural numbers and other inductively defined structures, together with suitable principles of definition by recursion. If we do that, the simply typed lambda calculus starts to look more like a real programming language, and we can think of simple type theory of a way of extending it with support for writing formal specifications and verifying that they are met. Dependent type theory goes further along this path, augmenting the programming language with a more expressive type system and integrating it more closely with the specification language.

In short, a type theory can be viewed either as a proof system in which we can define computable functions if we want to, or as a programming language that allows us to state and prove theorems as well. It is mainly a historical artifact that implementations of simple type theory are usually based on classical logic and favor the first viewpoint, while implementations of dependent type theory are usually based on a constructive logic and favor the second. In this section we will emphasize the constructive viewpoint, in order to balance the classical viewpoint we have developed so far. But it is important to recognize that dependent type theory provides an expressive framework for carrying out classical mathematics as well, and we will describe the axioms and constructions that support that in Section~\ref{subsection:lf:dependent:type:theory:additional:axioms}.

Dependent type theory goes by many names, including \emph{constructive type theory} and \emph{intuitionistic type theory}. The next two subsections present the core syntax and basic type constructions, and the subsections that follow present a number of extensions and variations. To avoid introducing names for theories and relating them to existing provers in a piecemeal fashion, we summarize the state of affairs here. Most systems today implement versions of dependent type theory with Pi types, type universes, and inductive types, from which other type constructions can be defined. A key distinguishing feature is whether the system in question implements an impredicative type of propositions, as described in Section~\ref{subsection:lf:the:type:of:propositions}. Systems that do not are generally classified as versions of \emph{predicative type theory} or \emph{Martin-L\"of type theory}, and systems that do are generally classified as versions of \emph{impredicative type theory} or the \emph{calculus of constructions}. Among contemporary proof assistants, Nuprl and Agda implement predicative type theories, and Coq, PVS, Matita, and Lean implement impredicative ones. As will be explained in the next subsection, another distinguishing feature is whether the logical foundation treats equality of types \emph{intensionally} or \emph{extensionally}. Of the systems just mentioned, only Nuprl and PVS are based on type theories that are extensional in this sense.

\subsection{Syntax}
\label{subsection:lf:dependent:type:theory:syntax}

The syntax of dependent type theory is more subtle than that of set theory and simple type theory. In set theory, the sets of terms, formulas, and proofs are defined by successive inductive definitions, each independent of the next. In simple type theory, there are, similarly, three kinds of syntactic objects: types, terms, and proofs. In contrast, dependent type theory can be presented in such a way that there is only one syntactic category, that of \emph{expressions}. Every expression has a type. Some expressions \emph{are} types, some expressions are propositions, and some expressions are proofs, and it is the type of the expression that determines which is the case. The tradeoff is that the definition of the set of well-formed expressions and the relation $t : \alpha$ that expresses that $t$ has type $\alpha$ are more complicated than the syntactic definitions of first-order logic and simple type theory.

In dependent type theory, a term $t$ with variables $\vec x$ is judged to be well formed with respect to a \emph{context} $\Gamma$ of the form $x_0 : \alpha_0{,}\; \ldots{,}\; x_{n-1} : \alpha_{n-1}$ that specifies the types of the variables. As with natural deduction and simple type theory, we can use the notation $\Gamma{,}\; x : \alpha$ to denote the extension of $\Gamma$ with the additional hypothesis $x : \alpha$, and we can use $\Gamma{,}\; \Delta$ for the union of two contexts. But now there is an important difference in that the order of the hypotheses matters. In the context $x_0 : \alpha_0{,}\; \ldots{,}\; x_{n-1} : \alpha_{n-1}$, a type expression $\alpha_j$ may depend on a variable $x_i$ for some $i < j$, and so exchanging $x_i$ and $x_j$ does not result in a well-formed context. The expression $\Gamma{,}\; \Delta$ should therefore be interpreted as describing a combination of the two contexts that preserves the dependency structure. Theoretical presentations of dependent type theory generally take a context to be a sequence, and they generally include explicit \emph{structural rules} that allow for weakening a context, contracting duplicate elements, and reordering elements subject to the dependency rules. In contrast, implementations of dependent type theory generally use efficient data structures and algorithms to manage and merge contexts.

The following three judgments are usually defined simultaneously with a mutual inductive definition:
\begin{itemize}
  \item the judgment that $\Gamma$ is a valid context;
  \item the judgment $\Gamma \fCenter t : \alpha$, which says that expression $t$ has type $\alpha$ in context $\Gamma$;
  \item the judgment $s \equiv t$, which says that expressions $s$ and $t$ are \emph{definitionally equal}.
\end{itemize}

In dependent type theory, propositions are represented as types, and if $\alpha$ is a proposition, the judgment $t : \alpha$ represents the assertion that $t$ is proof of $\alpha$. The most important task a proof assistant's kernel has to perform is therefore validating judgments of the form $\Gamma \fCenter t : \alpha$. For that purpose, a backward-style presentation of the rules is appropriate (Section~\ref{subsection:lf:first:order:logic:syntax}). To perform the task, type-checking algorithms also need to \emph{infer} the type of an expression; that is, given a well-formed expression $t$ in context $\Gamma$, they need to determine a type $\alpha$ such that $\Gamma \proves t : \alpha$. See Chapter~7 for details.


Let us start with an expression, $\Type$. For the moment we relinquish the requirement that every expression has a type, so we will not worry about assigning a type to $\Type$. (We will return to that issue in Section~\ref{subsection:lf:dependent:type:theory:universes}.) We will read judgments of the form $\Gamma \fCenter \alpha : \Type$ as ``$\alpha$ is a type in context $\Gamma$.'' We add the following two rules for generating contexts:
\begin{itemize}
  \item The empty context is a context.
  \item If $\Gamma$ is a context, $\alpha$ is a type in context $\Gamma$, and $x$ is a fresh variable, then $\Gamma{,}\; x : \alpha$ is a context.
\end{itemize}
In analogy to the assumption rule in natural deduction, we have:
\begin{itemize}
  \item If $x : \alpha$ is in $\Gamma$, then $\Gamma \fCenter x : \alpha$.
\end{itemize}

We also specify that definitional equality is reflexive, symmetric, and transitive, and that it respects the typing judgments: if $\Gamma \proves t : \alpha$ and we have $t \equiv t'$ and $\alpha \equiv \alpha'$, then $\Gamma \proves t' : \alpha'$. In the next subsection, we will see why such a rule is necessary. Definitional equality is also called \emph{judgmental equality}, and sometimes \emph{computational equality}, because it can usually be determined in principle by reducing terms to a normal form. The notion of $\beta$-equivalence for terms in the simply typed lambda calculus is a prototypical example of a notion of definitional equality.

Saying that two expressions $t$ and $t'$ are definitionally equal says that the logical foundation treats them as being essentially the same. This stands in contrast to \emph{propositional equality}, namely, judgments of the form $\Gamma \proves p : s =_\alpha t$, where the type $s =_\alpha t$ represents the \emph{proposition} that the terms $s$ and $t$ of type $\alpha$ are equal. Expressions that are definitionally equal are propositionally equal, by the reflexivity of propositional equality. But the converse does not hold in general: since all the resources of mathematics can be brought to bear to prove that two expressions are equal, one cannot identify propositional equality with definitional equality without making type checking undecidable.

\emph{Extensional} versions of type theory do add a rule saying that if two expressions are propositionally equal, they are definitionally equal. As a result, type checking in such systems sometimes raises proof obligations for users to meet. Although this section focuses on \emph{intensional} type theory (that is, type theory without that additional rule), almost everything said below applies to both kinds of systems.

\subsection{Basic type constructions}
\label{subsection:lf:basic:type:constructions}

Most rules in dependent type theory are associated with a single type construction and fall into one of the following categories:
\begin{itemize}
\item \emph{type formation rules}, resulting in a judgment of the form $\Gamma \fCenter \alpha : \Type$, where $\alpha$ is an instance of the new type construction;
\item \emph{introduction rules}, which show how to derive a judgment $\Gamma \fCenter t : \alpha$;
\item \emph{elimination rules}, which show how to make use of a hypothesis $x : \alpha$ in the context;
\item \emph{conversion rules}, which declare definitional identities, for example asserting that an introduction followed by an elimination can be simplified.
\end{itemize}
We will follow this organizational scheme where possible.


In dependent type theory, the function types $\alpha \to \beta$ are generalized by the \emph{Pi types}, $\PiT {x : \alpha} \beta$, in which the variable $x$ is bound. An element of this type should be viewed as a function $f$ that maps any element $x : \alpha$ to an element of type $\beta$, where $\beta$ can depend on $x$. Such a function is therefore called a \emph{dependent function}. The formation, introduction, and elimination rules are as follows:\strut
\begin{center}
\AXN{\strut\Gamma \fCenter \alpha : \Type}
\AXN{\strut\Gamma{,}\; x : \alpha \fCenter \beta : \Type}
\BIN{\strut\Gamma \fCenter (\PiT {x :\alpha} \beta) : \Type}
\DP \\
\medskip
\AXM{\strut\Gamma{,}\; x : \alpha \fCenter t : \beta}
\UIM{\strut\Gamma \fCenter (\Lam {x : \alpha} t) : (\PiT {x : \alpha} \beta)}
\DP
\quad\quad
\AXN{\strut\Gamma \fCenter t : (\PiT {x : \alpha} \beta)}
\AXN{\strut\Gamma \fCenter a : \alpha}
\BIN{\strut\Gamma \fCenter t \, a : \beta[a/x]}
\DP
\end{center}
The conversion rule is $(\Lam x t) \, s \equiv t[s/x]$. The usual construction $\alpha \to \beta$ represents the special case in which $\beta$ does not depend on $x$.

\emph{Sigma types} are another fundamental construction. An element $p$ of $\SigmaT {x : \alpha} \beta$ should be viewed as a pair $(a, b)$ where $a : \alpha$ and $b : \beta[a/x]$. In other words, the Sigma type is a generalization of the cartesian product $\alpha \times \beta$ with elements $(a, b)$ that are \emph{dependent pairs}, in the sense that type of the second component can depend on the first. The formation, introduction, and elimination rules are as follows:\strut
\begin{center}
\AXN{\strut\Gamma \fCenter \alpha : \Type}
\AXN{\strut\Gamma{,}\; x : \alpha \fCenter \beta : \Type}
\BIN{\strut\Gamma \fCenter (\SigmaT {x :\alpha} \beta) : \Type}
\DP
\quad\quad
\AXN{\strut\Gamma \fCenter a : \alpha}
\AXN{\strut\Gamma \fCenter b : \beta[a/x]}
\BIN{\strut\Gamma \fCenter (a, b) : (\SigmaT {x : \alpha} \beta)}
\DP \\
\medskip
\AXN{\strut\Gamma{,}\; z : (\SigmaT {x : \alpha} \beta) \fCenter \gamma : \Type}
\AXN{\strut\Gamma{,}\; x : \alpha{,}\; y : \beta \fCenter t : \gamma[(x, y) / z]}
\AXN{\strut\Gamma \fCenter p : (\SigmaT {x : \alpha} \beta)}
\TIN{\strut\Gamma \fCenter \fn{cases}_{\alpha, \Lam x \beta, \Lam z \gamma} \, (\Lam {x, y} t) \, p : \gamma[p / z]}
\DP
\end{center}
The elimination rule, which is analogous to the elimination rule for the existential quantifier in natural deduction, should be understood as follows. Suppose $\gamma$ is a type depending on a value $z$ of type $\SigmaT {x : \alpha} \beta$. Suppose also that $t$ is an expression that, given $x : \alpha$ and $y : \beta$, returns a value of type $\gamma[(x, y)/z]$. Then for any $p$ of type $\SigmaT {x : \alpha} \beta$, assuming $p$ represents the term $(a, b)$, the expression $\fn{cases} \, (\Lam {x, y} t) \, p$ is supposed to be represent $t [a /x, b / y] : \gamma[a/x, b/y]$. The conversion rule says exactly that:
\[
  \fn{cases} \, (\Lam {x, y} t) \, (a, b) \; \equiv \; t [a /x, b / y]
\]
We will see in Section~\ref{subsection:lf:dependent:type:theory:inductive:types} that a Sigma type is an instance of an inductively defined type. If we think of the introduction rule as describing the canonical way to build an element of type $\SigmaT {x : \alpha} \beta$, the elimination rule says that every element of $\SigmaT {x : \alpha} \beta$ arises in such a way.

A Pi type $\PiT {x : \alpha} \beta$ is analogous to a set-theoretic product $\prod_{x \in A} B_x$, and a Sigma type $\SigmaT {x : \alpha} \beta$ is analogous to set set-theoretic disjoint sum $\sum_{x \in A} B_x$. This explains the use of $\Pi$ and $\Sigma$ in the notation, as well as the fact that Pi types and Sigma types are sometimes called \emph{dependent products} and \emph{dependent sums}, respectively. But be careful: some authors use the phrase \emph{dependent product} to refer to Sigma types instead of Pi types, because the Sigma types can also be viewed as generalizations of the binary cartesian product.

Using $\fn{cases}$, we can define the projections:
\begin{align*}
 & \fn{fst} : (\SigmaT {x : \alpha} \beta) \to \alpha
&& \fn{snd} : (\PiT {z : (\SigmaT {x : \alpha} \beta)} {\beta[\fn{fst} \, z / x])}
\end{align*}
Given $a : \alpha$ and $b : \beta[a / x]$, these satisfy $\fn{fst} \, (a, b) \equiv a$ and $\fn{snd} \, (a, b) \equiv b$. The second of these identities illustrates the necessity of allowing definitional equality in type checking, since the left-hand side has type $\beta[\fn{fst} \, (a, b) / x]$ and the right-hand side has type $\beta[a / x]$. A type checker has to reduce $\fn{fst} \, (a, b)$ to $a$ in order to recognize that these types are the same.

In the elimination rule, the lambdas in the expression $\fn{cases}_{\alpha, \Lam x \beta, \Lam z \gamma} \, (\Lam {x, y} t) \, p$ are meant to indicate that the associated variables are bound by the construction. With the availability of Pi types, we could instead introduce $\fn{cases}_{\alpha, \Lam x \beta, \Lam z \gamma}$ as an expression on its own, with the following type:
\[
  (\PiT {x : \alpha, y : \beta} {\gamma[(x, y) / z]}) \to \PiT {z : (\SigmaT {x : \alpha} \beta)} \gamma
\]
In Section~\ref{subsection:lf:dependent:type:theory:universes}, we will explain how $\Type$ itself can be treated as a type, at which point $\fn{cases}$ itself can be viewed as a constant of the following type:
\begin{align*}
  & \PiT {\alpha : \Type{,}\; \beta : \alpha \to \Type{,}\; \gamma : (\SigmaT {x : \alpha} \beta) \to \Type} {} \\[-\jot]
  & \quad (\PiT {x : \alpha, y : \beta \, x} {\gamma \, (x, y)}) \to \PiT {z : (\SigmaT {x : \alpha} \beta)} {\gamma \, z}
\end{align*}
Using constants instead of constructions in this way is analogous to using axioms instead of rules in a presentation of first-order logic. Such approaches are generally favored in implementations, since it is easy to add new constants once the system can handle Pi types appropriately. We will continue to favor a rule-based presentation, however, since that usually clarifies the main ideas.

In a similar manner, we can go on to introduce the following types:
\begin{itemize}
\item sum types, $\alpha + \beta$;
\item the natural numbers, $\bN$;
\item the type $\ty{Empty}$ with no elements;
\item the type $\ty{Unit}$ with a single element;
\item the type $\ty{Bool}$ with two elements.
\end{itemize}
Sum types, $\bN$, and $\ty{Bool}$ were described in Section~\ref{subsection:lf:simply:typed:lambda:calculus:extensions}. For the type $\ty{Empty}$, there is no introduction rule, and the elimination rule says that we can map an element of empty to any type (which should be vacuously true because there are no elements of that type). The types $\ty{Empty}$, $\ty{Unit}$, and $\ty{Bool}$ are sometimes written $\mathbf 0$, $\mathbf 1$, and $\mathbf 2$, respectively, with the happy consequence that $\mathbf 2$ is isomorphic to $\mathbf 1 + \mathbf 1$. The rules follow the same patterns as the rules for Pi types and Sigma types; the key difference from the simply typed lambda calculus is that the types in the elimination rules can depend on the variable being eliminated. For example, the formation and introduction rules for $\bN$ are as follows:\strut
\begin{center}
  \AXN{\strut\Gamma \fCenter \bN : \Type}
  \DP
  \quad\quad
  \AXN{\strut\Gamma \fCenter 0 : \bN}
  \DP
  \quad \quad
  \AXN{\strut\Gamma{,}\; x : \bN \fCenter \fn{succ} \, x : \bN}
  \DP
\end{center}
Here is the elimination rule:
\begin{center}
  \AXN{\strut\Gamma{,}\; z : \bN \fCenter \gamma : \Type}
  \AXN{\strut\Gamma \fCenter s : \gamma(0)}
  \AXN{\strut\Gamma{,}\; x : \bN{,}\; y : \gamma(x) \fCenter t : \gamma(\fn{succ} \, x)}
  \AXN{\strut\Gamma \fCenter z : \bN}
  \QIN{\strut\Gamma \fCenter \fn{rec}_{\Lam z \gamma} \, s \, (\Lam {x, y} t) \, z : \gamma}
  \DP
\end{center}
For brevity, we have written $\gamma(u)$ for $\gamma[u/z]$. There are two conversion rules:
\begin{align*}
 & \fn{rec} \, s \, (\Lam {x, y} t) \, 0 \; \equiv \; s
&&  \fn{rec} \, s \, (\Lam {x, y} t) \, (\fn{succ} \, u) \; \equiv \; t[u/x, \fn{rec} \, s \, (\Lam {x, y} t) \, u / y]
\end{align*}
With the mechanism introduced in Section~\ref{subsection:lf:dependent:type:theory:universes} to abstract over types, we can express $\fn{rec}$ as a constant with the following type:
\[
\PiT {\gamma : (\bN \to \Type)} {\gamma \, 0 \to (\PiT {x : \bN} {\gamma \, x \to \gamma \, (\fn{succ} \, x)}) \to \PiT {x : \bN} {\gamma \, x} }
\]
Notice that this is essentially the second-order formulation of the induction axiom, with $\Prop$ replaced by $\Type$.

\subsection{Using propositions as types}
\label{subsection:lf:using:propositions:as:types}

The rules formulated in the previous section allow for the fact that a type $\alpha$ can depend on a variable $x$, but we have not yet provided any way of introducing such a dependency. To do so, we will invoke the propositions-as-types interpretation, introduced in Section~\ref{subsection:lf:simply:typed:lambda:calculus:semantics}.

Remember that, according to the interpretation, we view propositions as data types and proofs of those propositions as terms of those types. Pi and Sigma types can be interpreted in that light. If $\alpha$ is a type that depends on $x$ and we view $\alpha$ as a proposition about $x$, then instead of interpreting a term $p$ of type $\PiT x \alpha$ as a function, we should interpret it as a proof of $\Forall x \alpha$, namely, a recipe that, for every $x$, constructs a proof of $\alpha(x)$. Similarly, a term $p$ of type $\SigmaT x \alpha$ should be viewed as a constructive proof of $\Exists x \alpha$, providing a witness $x$ and a proof of $\alpha(x)$. Thus $\Pi$ and $\Sigma$ are simply the constructive interpretations of $\forall$ and $\exists$. The function type construction $\alpha \Implies \beta$ is identified with implication: constructively, a proof of $\alpha \Implies \beta$ is a recipe that transforms a proof of $\alpha$ into a proof of $\beta$, or a recipe that transforms evidence for $\alpha$ into evidence for $\beta$.

In first-order logic, given any terms $a$ and $b$, we can express the proposition that $a$ is equal to $b$. To do this in type theory, we simply add a new construction, the \emph{identity types}. Given $a : \alpha$ and $b : \alpha$, we introduce a type $\ty{Id}_\alpha \, a \, b$, which we will write $a =_\alpha b$. Drawing on propositions as types, we should think of an element $p : a =_\alpha b$ as a proof, or some sort of evidence, that $a$ and $b$ are equal. The formation and introduction rules are as follows:\strut
\begin{center}
  \AXN{\strut\Gamma \fCenter \alpha : \Type}
  \AXN{\strut\Gamma \fCenter a : \alpha}
  \AXN{\strut\Gamma \fCenter b : \alpha}
  \TIN{\strut\Gamma \fCenter a =_\alpha b : \Type}
  \DP
  \quad\quad
  \AXN{\strut\Gamma \fCenter \alpha : \Type}
  \AXN{\strut\Gamma \fCenter a : \alpha}
  \BIN{\strut\Gamma \fCenter \fn{refl}_\alpha \, a : a =_\alpha a}
  \DP
\end{center}
The elimination rule is as follows: given
\begin{itemize}
  \item $\Gamma{,}\; x : \alpha{,}\; y : \alpha{,}\; p : x =_\alpha y \fCenter \gamma : \Type$,
  \item $\Gamma{,}\; z : \alpha \fCenter t : \gamma[z/x, z/y, \fn{refl}_\alpha \, z/ p]$,
  \item $\Gamma \fCenter a : \alpha$,
  \item $\Gamma \fCenter b : \alpha$,
  \item $\Gamma \fCenter u : a =_\alpha b$,
\end{itemize}
we have $\fn{cases}_{\alpha, \Lam {x, y, p} {\gamma}} \, (\Lam z t) \, a \, b \, u : \Gamma[a/x, b/y, u/p]$. (We should really write $\fn{cases}^=$ instead of $\fn{cases}$ to distinguish it from the $\fn{cases}$ functions for Sigma types and sum types, but we will leave such distinctions implicit.) The rule says that if we have $u : a =_\alpha b$, we can, in a sense, assume that $u$ is really $\fn{refl} \, a : a =_\alpha a$. For brevity we will omit the conversion rule which says, roughly, that casing on reflexivity is the identity operation.

The elimination rule is a generalization of the substitution rule for equality, which asserts that given $e : a =_\alpha b$ and $h : \gamma \, a$ we can conclude $\fn{subst} \, e \, h : \gamma \, b$. If we think of $\gamma$ as a parameterized data type instead of a proposition, the substitution rule has a novel interpretation. Given an object $h$ of type $\gamma \, a$ and the fact $e : a =_\alpha b$, the types $\gamma \, a$ and $\gamma \, b$ should be equal, and so $h$ should have type $\gamma \, b$ as well. But unless $\gamma \, a$ and $\gamma \, b$ are definitionally equal, the type checker cannot tell that they are the same type. The additional information $e$ is needed to help the type checker see that they are, and the expression $\fn{subst} \, e \, h$ therefore serves to \emph{cast} or \emph{transport} $h$ along the equality to an object of type $\gamma \, b$.

With identity types in hand, we are off and running: we now have expressive means to define complex data types as well as state complex mathematical propositions. The propositions-as-types interpretation takes some getting used to, because types can be viewed as both data types and propositions, often at the same time. Viewed as a data type, the recursor for the natural numbers expresses the principle of recursive definition. Viewed as a proposition, it expresses the principle of induction. We can therefore use the recursor to define addition and multiplication, and then use the same recursor to prove things about them. For example, the type $\PiT {x, y : \bN} {x + y = y + x}$ expresses that addition is commutative, and a term of that type is interpreted as a proof of that fact.

Given $x$ and $y$ of type $\bN$, we can define $x \le y$ to be the type $\SigmaT {z : \bN} {x + z = y}$. In a similar way, we can define types $\ty{Divides} \, x \, y$, $\ty{Even} \, x$, and $\ty{Prime} \, x$. We can view an element of the type
\[
  \PiT {x : \bN} {\SigmaT {y : \bN} {(x \le y) \, \times \, \ty{Prime} \, y}}
\]
as a proof that for every $x$ there is a $y \ge x$ such that $y$ is prime. But we can also view it as a function which, given any $x$, returns such a prime, together with evidence that it meets this specification. For another example, we can express the Goldbach conjecture with the following type:
\[
  \PiT {x : \bN} {\ty{Even} \, x \; \times \; (x > 2) \Implies \SigmaT {y, z : \bN} {\ty{Prime} \, y \; \times \; \ty{Prime} \, z \; \times \; (x = y + z)}}
\]
A proof then includes a function that, given $x$ and evidence that $x$ is even, computes the relevant decomposition.

We can now see how dependent type theory allows us to do some of the things that were described in Section~\ref{subsection:lf:limitations:of:simple:type:theory} as falling outside the range of simple type theory. For example, given $n : \bN$, we can define the type $\ty{Fin} \; n$ to be $\SigmaT {x : \bN} {x < n}$, so that $\ty{Fin} \; n$ represents a type with $n$ elements. We can then define $\ty{Vec}_\alpha \; n$ to be $\ty{Fin} \; n \to \alpha$ and $\fn{Matrix}_\alpha \; m \, n$ to be $\ty{Fin} \; m \to \ty{Fin} \; n \to \alpha$.

\subsection{Inductive types}
\label{subsection:lf:dependent:type:theory:inductive:types}

We were able to define the natural numbers in set theory and simple type theory from axioms asserting the existence of an infinite set or type, respectively. We could try to do something similar in dependent type theory, but the rules we have introduced so far are not strong enough to carry out an impredicative construction. We could define the natural numbers using an axiom of infinity and the impredicative type of propositions described in Section~\ref{subsection:lf:the:type:of:propositions}, but the resulting type would not have nice computational properties.

In dependent type theory, it is more natural to add such inductive constructions axiomatically, with new type constructors and rules. This encourages the inductive constructions to be thought of abstractly, in terms of their constructors and recursion principles. For example, the type $\ty{List} \; \alpha$ has constructors $\fn{nil}_\alpha : \ty{List} \; \alpha$ and $\fn{cons}_\alpha : \alpha \to \ty{List} \; \alpha \to \alpha$. The recursor expresses a principle of recursion on lists, and the conversion rule says that the recursor satisfies the expected defining equations. In the next subsection we will see that once we have variables ranging over types, we can promote the subscripted $\alpha$ in $\fn{cons}_\alpha$ to a bona fide argument to $\fn{cons}$, resulting in a single $\fn{cons}$ function that is  \emph{polymorphic} over types.

The \emph{W Types}, introduced by Martin-L\"of, are a canonical family of inductively defined types. The formation and introduction rules are as follows:\strut
\begin{center}
  \AXN{\strut\Gamma \fCenter \alpha : \Type}
  \AXN{\strut\Gamma{,}\; x : \alpha \fCenter \beta : \Type}
  \BIN{\strut\Gamma \fCenter (\GenericBinder W {x : \alpha} \beta) : \Type}
  \DP \\
  \medskip
  \AXN{\strut\Gamma \fCenter a : \alpha}
  \AXN{\strut\Gamma{,}\; y : \beta[a/x] \fCenter f : (\GenericBinder W {x : \alpha} \beta)}
  \BIN{\strut\Gamma \fCenter \fn{sup}_{\alpha, \Lam x \beta} \, a \, f : (\GenericBinder W {x : \alpha} \beta)}
  \DP
\end{center}
An element of $\GenericBinder W {x : \alpha} \beta$ is intended to denote a well-founded tree whose nodes are labeled by elements of $\alpha$, with the property that the subtrees of a node labeled $a$ are indexed by the elements of $\beta[a/x]$. The tree $\fn{sup} \, a \, f$ is the tree whose top node is labeled $a$ and whose children are given by $f$. (In order for us to construct an element of this type, there has to be at least one $a$ in $\alpha$ such that $\beta[a/x]$ is empty, so that the single node labeled $a$ is a tree.) The elimination principle provides a recursion for such well-founded trees,  as follows. Given a family of types $\gamma$ indexed by trees,
\begin{center}
\AXN{\strut\Gamma{,}\; z : (\GenericBinder W {x : \alpha} \beta) \fCenter \gamma : \Type,}
\DP
\end{center}
and given a term $t$ that assigns a value to any tree of the form $\fn{sup} \, a \, f$ from an assignment $g$ of values to its subtrees,
\begin{center}
\AXN{\strut\Gamma{,}\; a : \alpha{,}\; f : (\beta[a/x] \to \GenericBinder W {x : \alpha} \beta), g : (\PiT {y : \beta[a/x]} \gamma[f \, y / z]) \fCenter t : \gamma [\fn{sup} \, a \, f / z],}
\DP
\end{center}
we have a value for every tree $u$,
\begin{center}
\AXN{\Gamma{,}\; u : (\GenericBinder W {x : \alpha} \beta) \fCenter \fn{rec}_{\Lam z \gamma} \, (\Lam {a, f, g} t) \, u : \gamma[u/z].}
\DP
\end{center}
The conversion rule, not shown, says that the recursor satisfies the expected defining equations.

Sigma types, cartesian products, and sum types can all be viewed as instances of inductive types. In fact, with the notion of \emph{indexed families of types}, even the identity types introduced in the last subsection become an instance of the general schema. For details, see Chapters~3 and 4.

\subsection{Universes}
\label{subsection:lf:dependent:type:theory:universes}

At this point, we can construct concrete types like $\bN \to \bN$, $\fn{List} \; \bN$, and $\ty{Fin} \; n$, but we do not have variables ranging over types. Recall that the rule for introducing a new variable into the context is as follows:
\begin{itemize}
\item If $\Gamma$ is a context and $\Gamma \fCenter \alpha : \Type$, then $\Gamma, \, x : \alpha$ is a context.
\end{itemize}
So, as things stand, we need a judgment $\Type : \Type$ in order to have a variable range over types. Unfortunately, this makes the system inconsistent \cite{girard:72}. A solution is to replace $\Type$ by a sequence of type universes, $\Type_0, \Type_1, \Type_2, \ldots$ with $\Type_i : \Type_{i+1}$. Now any judgment of the form $\alpha : \Type_i$ is interpreted as saying that $\alpha$ is a type, but every type belongs to a particular type universe. Sometimes the letter $\ty{U}$ is used to denote type universes instead, as in $\ty U_0, \ty U_1, \ty U_2, \dotsc$.

Once we have variables ranging over types, we can construct Pi types and lambda abstractions that bind those variables. Type formation rules need to specify which universe the constructions end up in. For Pi types, the formation rules becomes the following:\strut
\begin{center}
  \AXN{\strut\Gamma \fCenter \alpha : \Type_i}
  \AXN{\strut\Gamma{,}\; x : \alpha \fCenter \beta : \Type_j}
  \BIN{\strut\Gamma \fCenter (\PiT {x :\alpha} \beta) : \Type_{\fn{max} \, i \, j}}
  \DP
\end{center}
In other words, a Pi type lands in the larger of the two universes. An an example, the expression $\Lam {\tau : \Type_i} {\Lam {x : \tau} x}$, which represents the polymorphic identity function for types in the universe $\Type_i$, has type $\PiT {\tau : \Type_i} {\tau \to \tau}$. That expression, in turn, has type $\Type_{i + 1}$. The formation rules for inductive types also have to be adjusted accordingly.

Implementations of type theory with universes tend to use variables $i, j, \ldots$ ranging over the levels of a type universe so that definitions and theorems can be expressed generically and work when instantiated to any particular universe. This is known as \emph{universe polymorphism}, and it allows us to write $\ty{Type}$ instead of $\ty{Type}_i$ and leave it to the system to work out the constraints on the choice of $i$. This is often quite effective in providing the illusion that there is just one big universe of types. Some formulations of type theory add the following rule, for any $j > i$:\strut
\begin{center}
    \AXN{\strut\Gamma \fCenter \alpha : \Type_i}
    \UIN{\strut\Gamma \fCenter \alpha : \Type_j}
    \DP
\end{center}
This effectively makes the type universes \emph{cumulative}.

The ability to quantify over types means that we can make structures into first-class objects. For example, we can define the type $\ty{Group}_i$ of groups with carrier in the universe $\Type_i$ as a big Sigma type:
\begin{align*}
  & \SigmaT {\,  \ty{Carrier} : \Type_i} {} \\[-\jot]
  & \quad \SigmaT {\, \fn{op} : \ty{Carrier} \to \ty{Carrier} \to \ty{Carrier}{,}\; \fn{id} : \ty{Carrier}{,}\; \fn{inv} : \ty{Carrier} \to \ty{Carrier}} {} \\[-\jot]
  & \quad (\PiT {x, y, z : \ty{Carrier}} {\fn{op} \, (\fn{op} \, x \, y) \, z = \fn{op} \, x \, (\fn{op} \, y \, z)}) \times (\PiT {x : \ty{Carrier}} {\fn{op} \, x \, \fn{id} = x}) \times \cdots
\end{align*}
In words, a group consists of a type, $\ty{Carrier}$, a binary operation $\fn{op}$ on $\ty{Carrier}$, an identity element $\fn{id}$, and an inverse function $\fn{inv}$, satisfying the group axioms. It can take time to get used to the fact that the data associated with a group and its axiomatic properties are treated uniformly as elements of the structure, but this is a hallmark of dependent type theory, and it can be very useful.

We have now realized most of the design requirements enumerated in Section~\ref{subsection:lf:limitations:of:simple:type:theory}. Types are expressions, every expression has a type, and we have all the means of dependent type theory at our disposal to define new types as expressions of type $\Type_i$ for some $i$.

Expressions in type theory can be verbose. The polymorphic $\fn{cons}$ function on lists has type $\PiT {\tau : \Type} {\tau \to \ty{List} \; \tau \to \ty{List} \; \tau}$, so that given $n : \bN$ and $\ell : \ty{List} \; \bN$, we have to write $\fn{cons} \, \bN \, n \, \ell$ for the result of adding $n$ to the beginning of $\ell$. Implementations of type theory usually allow the first argument to $\fn{cons}$ to be marked \emph{implicit}. This allows users write $\fn{cons} \, n \, \ell$ and leave it to the system to infer that the first argument is $\bN$ from the types of $n$ and $\ell$. This process is described in detail in Chapter~7.

In dependent type theory, most concrete type constructions, like the number systems, land in the smallest universe, $\Type_0$. It is only generic constructions involving types that require larger universes. A type whose elements are structures with carrier in $\Type_i$ is itself an element of $\Type_{i+1}$. In category theory, one sometimes wants to consider a type of \emph{small} categories, where the objects are elements of a type in $\Type_{i+1}$, while the morphisms between any two objects are represented by a type in $\Type_i$. (For example, the type of groups with carrier in $\Type_i$ itself has type $\Type_{i+1}$, but the type of morphisms between any two such groups has type $\Type_i$.) The type of such small categories itself has type $\Type_{i+2}$.

\subsection{The type of propositions}
\label{subsection:lf:the:type:of:propositions}

The propositions-as-types interpretation, as we have developed it so far, highlights the parallels between data type constructions and logical operations. It reminds us that every mathematical statement has potential computational significance, whether or not we choose to think of it in those terms. But the conflation of propositions and types is sometimes unnatural. Elements of the types $\ty{Fin} \; x$, which were defined as $\SigmaT {y : \bN} {y < x}$ in Section~\ref{subsection:lf:using:propositions:as:types}, are best viewed as a piece of data meeting a specification. Describing them as constructive proofs that there exists something less than $x$ or as the combination of two pieces of data is awkward.

There are other reasons to resist conflating proofs and data. Suppose $(y, h)$ and $(y', h')$ are both elements of $\ty{Fin} \; x$. Under what conditions should we say that they are equal? Under a strict propositions-as-types interpretation, it is not enough to know $y = y'$; we also have to worry about the equality of $h$ and $h'$ under the cast that identifies their types. From a mathematical perspective, these proofs are irrelevant.

Another problem is that under the strict propositions-as-types interpretation, propositions are stratified by the type universes. A theorem about groups in type universe $i$, $\PiT {G : \ty{Group}_i} \dotsc$, is at best an element of $\ty{Type_{i+1}}$. We can no longer take a subset of a type $\alpha$ to be a function of type $\alpha \to \Prop$; rather, for each $i$, we can talk about the subsets of type $\alpha \to \Type_i$. So we cannot define the subgroup of a group $G$ generated by a set of elements $S$ to be the intersection of all the subgroups of $G$ containing $S$. Instead, given $\ty{Carrier} \; G : \Type_i$ and $S : \ty{Carrier} \; G \to \Type_j$, for any $k$ we can consider the sets of type $\ty{Carrier} \; G \to \Type_k$ containing $S$ and closed under the group operations. A universe calculation shows that the intersection of these subgroups has type $\ty{Carrier} \; G \to \Type_\ell$, where $\ell = \max (i, j, k+1)$. In particular, $\ell > k$, so the subgroup just defined is not among the subgroups in the intersection.

In other words, propositions as types, as we have understood it so far, serves to stratify sets into a hierarchy whereby any definition of a set that quantifies over a fixed totality of sets produces a new set that is not part of that totality. Such a treatment of sets and types is said to be \emph{predicative}, a term introduced early in the twentieth century in the wake of the set-theoretic paradoxes. Russell and Whitehead's original \emph{ramified type theory} had this character, but it was deemed unworkable for ordinary mathematics without an additional axiom, the \emph{axiom of reducibility}, which provided an ad hoc workaround.

A better solution is to introduce a type $\Prop$ like the one in simple type theory, with the property that a Pi type $\PiT {x : \alpha} \beta$ is in $\Prop$ whenever $\beta$ is in $\Prop$, no matter what type universe $\alpha$ is in. It is common to use the notation $\Forall {x : \alpha} \beta$ for such a Pi type, and putting it in $\Prop$ means that we can form a proposition by quantifying over the elements of any type. Such a type $\Prop$ is said to be \emph{impredicative}. One can think of a type $\alpha : \Prop$ is being inhabited if and only if $\alpha$ is true, in which case, a judgment $h : \alpha$ shows that it is true. The type $\Prop$ can be placed at the bottom of the hierarchy of universes by declaring $\Prop : \Type_0$, or alongside the smallest universe of data by declaring $\Prop : \Type_1$.

With such a type of propositions, one can use a general framework for inductive definitions to define logical operations. The connectives $\exists$, $\wedge$, $\vee$, and $\bot$ are defined just like $\Sigma$, $\times$, $+$, and $\ty{Empty}$, except that the results land in $\Prop$ rather than $\Type$. Identity types are also generally assumed to take values in $\Prop$. To balance the strength of impredicativity, restrictions are placed on the elimination rules. Saying that a type $\alpha$ is a proposition is meant to distinguish it from a data type; the existence of an element $h : \alpha$ is supposed to show that $\alpha$ is true but the object that $h$ denotes need not bear any additional information. The distinction means that, given $\alpha : \Type_i$ for some $i$, and given $\beta : \Prop$, there are two versions of the Sigma type construction $\SigmaT {x : \alpha} \beta$, depending on whether we view it as a data type or a proposition:
\begin{itemize}
  \item The \emph{subtype} $\{ x : \alpha \mid \beta \} : \Type_i$ consists of pairs $(a, h)$, where $a : \alpha$ and $h : \beta[a / x]$. The first component is data, and can be used in functions with any return type. The second component is a proposition.
  \item The proposition $(\Exists {x : \alpha} \beta) : \Prop$ says that there exists an $x$ satisfying $\beta$. An element $h$ of this type only guarantees the existence of such an $x$, but it does not make one available as data.
\end{itemize}

Classically, we can think of every element of $\Prop$ as being either a type with no elements, $\False$, or a type, $\True$, with a single element, $\fn{trivial} : \True$. In other words, classically, we can identify $\Prop$ with $\ty{Bool}$. We can formalize the statement that $\Prop$ is two-valued with the following strong form of the law of the excluded middle, known as \emph{propositional decidability}:
\[
  \Forall {\alpha : \Prop} {\alpha + \neg \alpha}
\]
This suffices to define an isomorphism between $\Prop$ and $\ty{Bool}$.

The following principle, known as \emph{proof irrelevance}, says that any two proofs of a proposition are equal.
\[
\Forall {\alpha : \Prop{,}\; h_1 : \alpha{,}\; h_2 : \alpha} {h_1 = h_2}
\]
Given any predicate $P : \alpha \to \Prop$, proof irrelevance implies that any two elements of the subtype $\{ x : \alpha \mid P \, x \}$ are equal if their first components are equal; in other words, only the value matters, not the justification. Proof irrelevance is consistent with the classical view of $\Prop$ described in the last paragraph, but it is also reasonable from a constructive viewpoint in which the operations that land in $\Prop$ are viewed as erasing computational information. The principle can even be implemented as a conversion rule, in which case it is called \emph{definitional proof irrelevance}. The notion of proof irrelevance is not limited to the presence of $\Prop$: even in predicative type theory, one can add an axiom that says that any two elements of an identity type $a =_\alpha b$ are equal. This principle is known as \emph{axiom K}.

To sum up, the introduction of an impredicative type universe $\Prop$ offers the following benefits:
\begin{itemize}
\item It provides a means to separate propositions from data.
\item It allows for impredicative definitions.
\item It provides a means of incorporating proof irrelevance.
\end{itemize}
In ordinary mathematics, the distinction between propositions and data is more natural.

\subsection{Additional axioms}
\label{subsection:lf:dependent:type:theory:additional:axioms}

The axiom of function extensionality, discussed in Section~\ref{subsection:lf:higher:order:logic}, can be extended to dependent functions:
\[
  \Forall {\alpha : \Type{,}\; \beta : \alpha \to \Type} \Forall {f, g : (\PiT {x : \alpha} \beta \, x)} {(\Forall {x : \alpha} {f \, x = g \, x}) \Implies f = g}
\]
The axiom of propositional extensionality is essentially unchanged:
\[
  \Forall {\alpha, \beta : \Prop} {(\alpha \Iff \beta) \Implies \alpha = \beta}.
\]
Neither of these is derivable from the axioms we have seen so far, but both can be consistently added.

In the context of dependent type theory with an impredicative type of propositions, there is an elegant formulation of the axiom of choice. Define $\ty{Nonempty} \; \alpha$ to be the proposition $\Exists {x : \alpha} \top$, which asserts that there exists an element of type $\alpha$. We can then assert the existence of a function that chooses an element from any nonempty type:
\[
  \fn{choice} : \PiT {\alpha : \Type} {\ty{Nonempty} \; \alpha \to \alpha}
\]
A version of the $\varepsilon$ function, discussed in Section~\ref{subsection:lf:formulations:of:simple:type:theory}, can be defined from that. The sketch of Diaconescu's theorem presented there
can be modified to derive propositional decidability from choice and extensionality.

The addition of propositional extensionality and function extensionality can have negative effects on computational reduction. Ideally, in a constructive type theory, every closed term of type $\bN$ should reduce to a numeral. The extensionality axioms, however, can block such reductions; for example, there is generally no way to reduce an expression $\fn{subst} \, e \, h$ when $e$ is an instance of the extensionality axiom, even though the substitution operator $\fn{subst}$ can be interpreted as a computationally inert operation that only serves to modify the type of $h$. But propositional and function extensionality are generally consistent with computational interpretations that ignore type information and replace $\fn{subst} \, e \, h$ by $h$. As a result, a constructive proof assistant that incorporates extensionality can still support code extraction. In contrast, the use of the choice axiom breaks the computational interpretation entirely. With functional extensionality, propositional extensionality, and choice, dependent type theory has a fully classical flavor.

Many other additions to type theory have been considered in the literature. For example, \emph{quotient types} \cite{DBLP:conf/tlca/Hofmann95,DBLP:conf/tphol/Nogin02} provide an axiomatic way of constructing quotients, and \emph{induction-induction} and \emph{induction-recursion} provide more general forms of inductively defined types. Many systems also implement \emph{coinductive types}, as described in Chapter~5.


\subsection{Homotopy type theory}
\label{subsection:lf:homotopy:type:theory}

\emph{Homotopy type theory} draws on the fact that predicative type theory has a homotopy-theoretic interpretation, first discovered by Steve Awodey and Michael Warren on the one hand, and Vladimir Voevodsky, independently, on the other. Rather than thinking of types as data types, we can think of them as topological spaces, and rather than think of terms as expressions that depend on the values of their variables, we can think of them as terms that depend \emph{continuously} on those variables. Even better, we can think of types as abstract representations of spaces up to the topological notion of \emph{homotopy equivalence}. Awodey and Warren made this precise by showing that dependent type theory can be interpreted in any \emph{model category} with some additional structure. Voevodsky interpreted it in the category of \emph{simplicial sets}, a structure that is fundamental to algebraic topology.

In the homotopy-theoretic interpretation, elements of an identity type $a =_\alpha b$ represent \emph{paths} from $a$ to $b$, that is, continuous maps of the unit interval into $\alpha$ that map one endpoint to $a$ and the other to $b$. The elimination rule for identity types then corresponds to a method of proof in algebraic topology that involves contracting paths down to a constant path. In type theory, given $p, q : a =_\alpha b$, one can ask whether there is an element of $p =_{a =_\alpha b} q$. With proof irrelevance or axiom K, the answer is trivially yes, but in the homotopy-theoretic interpretation, it amounts to the interesting question as to whether there is a continuous deformation between the paths $p$ and $q$. Whereas proof irrelevance trivializes questions about elements of the identity types, homotopy type theory makes them a central focus.

Given $\alpha, \beta : \Type$, one can formally define a type $\alpha \sim \beta$ of \emph{equivalences} between types, corresponding to the notion of \emph{homotopy equivalence} in the interpretation just described. From the axioms for equality, it is easy to construct an expression of type $\alpha = \beta \to \alpha \sim \beta$. Voevodsky's \emph{univalence axiom} asserts that this map is an equivalence, so, in particular, there is a map of type $\alpha \sim \beta \to \alpha = \beta$ going in the opposite direction. Voevodsky showed that it is consistent to add this principle to predicative type theory. Informally, this can be read as the statement that isomorphic types are equal, allowing us to use the laws of the equality to substitute one structure for another freely once we have constructed a bijection between them. The univalence axiom supports the informal mathematical practice of treating isomorphic structures as being the same; its consistency hinges on the fact that the things we can say about structures in the language of dependent type theory respect isomorphism. Read computationally, the univalence axiom says that once we have a translation between two data types, we may exchange them freely in a program.

When it was initially presented, the univalence axiom did not have a computational interpretation. A good deal of work in recent years has done toward developing foundations for homotopy type theory with better computational properties. In particular, versions of \emph{cubical type theory} incorporate the dependence of terms and types on special \emph{interval variables}. In such systems, the elimination rule for identity types and the univalence principle are derivable from more fundamental syntactic rules.

At present, there are a number of prototype systems based on cubical type theory, including the Arend proof assistant \cite{arend} and a cubical version of Agda \cite{DBLP:journals/pacmpl/VezzosiM019}. Such systems are not sufficiently well settled to give a definitive presentation now, but the underlying ideas offer promising new directions for interactive theorem proving.

\section{Bibliographical Notes}

Most of the information in this chapter can be found in numerous textbooks and other expository sources. The suggestions that follow are only meant to offer starting points for further reading.

For a broad history of logic, see Kneale and Kneale \cite{MR0144812}. For a history of logic from around the turn of the twentieth century, see Van Heijenoort \cite{MR0209111} or Mancosu, Zach, and Badesa \cite{MR2895611}. For important source documents in the foundations of mathematics, see Ewald \cite{ewald:96}. For the history of constructive mathematics, see Troelstra and Van Dalen \cite{MR966421:modified} or Beeson \cite{MR786465}. For a history of interactive theorem proving, see Harrison, Urban, and Wiedijk \cite{DBLP:series/hhl/HarrisonUW14}.

Most of the general background in logic discussed here is presented in greater detail in Avigad \cite{avigad:nd}. Countless textbooks cover the basics of first-order logic. The \emph{Handbook of Mathematical Logic} \cite{MR457132:modified} is a general reference, and the \emph{Handbook of Proof Theory} \cite{MR1640324} and Troelstra and Schwichtenberg \cite{MR1776976} offer proof-theoretic perspectives. For classical, model-theoretic semantics, see Chang and Keisler \cite{MR1059055}, Hodges \cite{MR1221741}, or Marker \cite{MR1924282}. For algebraic and category-theoretic approaches to semantics, including the semantics of higher-order logic, see Troelstra and Van Dalen \cite{MR966421:modified}, Lambek and Scott \cite{MR856915}, or Mac Lane and Moerdijk \cite{MR1300636}. Harrison \cite{DBLP:books/daglib/0022394} provides decision procedures for all the decidable theories enumerated in Section~\ref{subsection:lf:decidability:and:incompleteness}. Smith \cite{MR3059012} is an introduction to theories of arithmetic and the incompleteness theorems.

For more information on the simply typed lambda calculus and related topics, see Hindley and Seldin \cite{MR2435558}, Barendregt, Dekkers, and Statman \cite{MR3114769}, Girard \cite{MR1003608}, or S{\o}rensen and Urzyczyn \cite{sorensen:urzyczyn:06}. The last two offer introductions to the propositions-as-types interpretation, the history of which is surveyed by Wadler \cite{DBLP:journals/cacm/Wadler15}.

Good introductions to set theory include Devlin \cite{MR1237397}, Enderton \cite{MR0439636}, and Kunen \cite{MR2905394}. Grabowski, Kornilowicz, and Naumowicz \cite{DBLP:journals/jfrea/GrabowskiKN10} describe its implementation in Mizar. Higher-order logic is treated in Andrews \cite{MR1932484}, Takeuti \cite{MR882549}, or Lambek and Scott \cite{MR856915}, and it is informative to see how it is implemented in Isabelle \cite{DBLP:books/sp/NipkowPW02}, HOL4 \cite{HOL4}, and HOL Light \cite{hol:light:tutorial}. For introductions to dependent type theory, see Martin-L\"{o}f \cite{MR769301}, Nordstr\"{o}m, Petersson, and Smith \cite{MR1243882}, or Nederpelt and Geuvers \cite{MR3445957}. Once again, it is informative to see how the theory is put into practice in Coq \cite{DBLP:series/txtcs/BertotC04}, Agda \cite{DBLP:conf/afp/Norell08}, Nuprl \cite{DBLP:books/daglib/0068834}, PVS \cite{pvs}, and Lean \cite{theorem:proving:in:lean}. The \emph{Homotopy Type Theory} book \cite{DBLP:books/daglib/0046165} describes early systems of homotopy type theory, and its appendices provide a clear presentation of the rules of predicative type theory.

\newcommand{\acks}[1]{\bigskip \noindent \emph{Acknowledgements.} #1}

\acks{I am grateful to Guillaume Dubach, Randy Pollack, and Pedro S\'anchez Terraf for comments and corrections.}

\bibliographystyle{plain}
\bibliography{bib}

\end{document}